\documentclass[%
preprint,
superscriptaddress,
nofootinbib,
 amsmath,amssymb,
 aps,
]{revtex4-2}

\usepackage{todonotes}
\usepackage{graphicx,subfigure}% Include figure files
\usepackage{dcolumn}% Align table columns on decimal point
\usepackage{bm}% bold math
\usepackage[T1]{fontenc}
\usepackage[section]{placeins}
\RequirePackage{mathpazo} %changes font to Palatino
\usepackage{textcomp}
\usepackage{float}
\usepackage{braket}
\usepackage{amsmath}
\usepackage{mathtools}
\usepackage[normalem]{ulem}

\usepackage{hyperref}% add hypertext capabilities
\usepackage{xcolor}
\definecolor{red}{rgb}{0.9,0,0}
\definecolor{magenta}{rgb}{1.0,0,1.0}

\usepackage{enumitem,amssymb}
\newlist{todolist}{itemize}{2}
\setlist[todolist]{label=$\square$}
\usepackage{pifont}

\usepackage{makecell}
\begin{document}

%\setlength{\dblfloatsep}{10pt plus 2pt minus 2pt}

%\title{Robust Amplitude Estimation as an application benchmark for near-term quantum devices}
\title{Experimental demonstration of Robust Amplitude Estimation on near-term quantum devices for chemistry applications}

%%% Note: author order hasn't been discussed yet
\author{Alexander A. Kunitsa}
\email{aakunitsa@gmail.com}
\affiliation{Zapata Computing, Inc., 100 Federal St., Boston, MA 02110, USA}
\author{Nicole Bellonzi}
\affiliation{Zapata Computing, Inc., 100 Federal St., Boston, MA 02110, USA}
\author{Shangjie Guo}
\affiliation{Fuels \& Low Carbon Tech - Digital Science, bp Technology, 501 Westlake Park Blvd, Houston, TX 77079, USA}
\author{J\'er\^ome F. Gonthier}
\affiliation{Zapata Computing, Inc., 100 Federal St., Boston, MA 02110, USA}
\author{Corneliu Buda}
\affiliation{Technology LCP - Low Carbon Energy Advanced Bio \& Physical Science, 30S Wacker Drive, Chicago, IL 60606, USA}
\author{Clena M. Abuan}
\affiliation{Fuels \& Low Carbon Tech - Digital Science, bp Technology, 501 Westlake Park Blvd, Houston, TX 77079, USA}
\author{Jhonathan Romero}
\affiliation{Zapata Computing, Inc., 100 Federal St., Boston, MA 02110, USA}

\date {\today}

\begin{abstract}
This study explores hardware implementation of Robust Amplitude Estimation (RAE) on IBM quantum devices, demonstrating its application in quantum chemistry for one- and two-qubit Hamiltonian systems. Known for potentially offering quadratic speedups over traditional methods in estimating expectation values, RAE is evaluated under realistic noisy conditions.
Our experiments provide detailed insights into the practical challenges associated with RAE.
We achieved a significant reduction in sampling requirements compared to direct measurement techniques. 
In estimating the ground state energy of the hydrogen molecule, the RAE implementation demonstrated two orders of magnitude better accuracy for the two-qubit experiments and achieved chemical accuracy.
These findings reveal its potential to enhance computational efficiencies in quantum chemistry applications despite the inherent limitations posed by hardware noise.
We also found that its performance can be adversely impacted by coherent error and device stability and does not always correlate with the average gate error.
These results underscore the importance of adapting quantum computational methods to hardware specifics to realize their full potential in practical scenarios.
\end{abstract}

\maketitle

\section{Introduction}
\label{sec:intro}
Continuous progress in quantum hardware brings us closer to demonstrating the utility of quantum computing for industrial applications. 
Quantum computing holds the promise of revolutionizing fields such as chemistry, materials science, and optimization by solving problems intractable for classical computers~\cite{dalzell_quantum_2023}. 
Since modern devices are limited in the qubit count, coherence times, and gate fidelities, the near-term quantum algorithms must be applied in tandem with error mitigation techniques or incorporate noise models in their design. 

A prominent example of such algorithms is Robust Amplitude Estimation (RAE)~\cite{wang2020bayesian, ELF}, an enhanced sampling technique that uses short-depth quantum amplitude amplification~\cite{suzuki_amplitude_2020, tanaka_amplitude_2021} to accelerate observable estimation. 
It was initially developed to address the "measurement bottleneck"~\cite{gonthier_measurements_2022} plaguing variational algorithms, particularly the variational quantum eigensolver (VQE)~\cite{Peruzzo2014}. 
To that end, RAE can reduce the scaling of the number of state preparations needed to evaluate the VQE cost function with precision $\epsilon$ from $\mathcal{O}(\frac{1}{\epsilon^2})$ to $\mathcal{O}(\frac{1}{\epsilon^{\alpha}})$, where $\alpha \in [1, 2]$. 

While RAE alone is likely not enough to make VQE practical~\cite{johnson_reducing_2022}, there are many cases where the estimation of expectation values is an essential component of a quantum algorithm that can benefit from RAE. 
Examples include estimating magnetization in quantum materials~\cite{oftelie_simulating_2021}, simulation applications, the determination of properties over time-evolved solutions of differential equations~\cite{jaksch_variational_2022, leong_variational_2022}, or estimating Betti numbers in topological data analysis~\cite{hayakawa_quantum_2022,berry_analyzing_2024}. 
This motivated further research into RAE-related algorithms, such as  Quantum Amplitude Estimation (QAE)~\cite{suzuki_amplitude_2020} and their applications outside the chemistry domain~\cite{giurgica-tiron_low_2022, giurgica-tiron_low-depth_2022}.

The key feature of RAE is the ability to trade circuit depth to reduce the runtime. This flexibility comes at the cost of making the algorithm dependent on the device parameters, in particular, the gate fidelity~\cite{katabarwa_reducing_2021}. 
Further, it assumes the depolarizing noise model accurately captures the underlying decoherence mechanisms. If this assumption is not satisfied in experiments on quantum hardware, special techniques, such as randomized compiling, need to be applied to ensure the validity of the RAE inference protocol. 
As shown by Dalal et al.~\cite{dalal_noise_2023}, this is especially true in the presence of a coherent error severely limiting the scalability of RAE on near-term and possibly early fault-tolerant quantum devices. Therefore, it is critically important to assess the algorithm fitness for a particular application in a realistic setting. 

Research has shown that early variants of QAE, such as Maximum Likelihood Quantum Amplitude Estimation (MLQAE)~\cite{suzuki_amplitude_2020} and Iterative QAE (IQAE)~\cite{grinko_iterative_2021}, can offer practical advantages in NISQ environments by optimizing the balance between circuit depth and estimation accuracy~\cite{rao2020quantum}. 
These approaches help mitigate the errors introduced by quantum noise, allowing for more accurate results with fewer quantum resources. 
Furthermore, implementations of low-depth QAE algorithms on state-of-the-art quantum devices, such as trapped-ion quantum computers, have demonstrated significant improvements in accuracy when advanced techniques like the Chinese Remainder Theorem (CRT) and Maximum Likelihood Estimation (MLE) are employed~\cite{giurgica-tiron_low-depth_2022}. 
These studies emphasize the need to tailor quantum algorithms to the specific noise characteristics and capabilities of the hardware to achieve optimal performance.
In more recent research for reducing the quantum resources for observable estimation, Amplified Amplitude Estimation (AAE) has emerged utilizing prior knowledge about the system to transform the problem into one where a small deviation from a known value can be efficiently estimated~\cite{simon2024amplified}.

As most prior work on RAE and related techniques focused on optimizing circuit depth and mitigating noise in NISQ environments, the effects of these optimizations on the statistical properties of the estimation process have not been studied in detail, and the impact of specific quantum device characteristics such as noise profiles has been often overlooked. 
This may limit the practical applicability of these algorithms in real-world quantum computing tasks, where hardware imperfections are significant. In particular, it remains unclear how the convergence properties of RAE are affected by device parameter instability~\cite{Dasgupta2021} and to what extent it can mitigate device noise when applied to estimating the expectation values of multi-term qubit operators, such as chemical Hamiltonians. 

In this paper, we present an end-to-end implementation of RAE on a premium IBM device, {\it ibmq\_montreal}, for estimating the ground state energy of one- and two-qubit Hamiltonians representing a model chemical system, $H_2$. 
In contrast to the previous experimental demonstrations of RAE~\cite{katabarwa_reducing_2021, dalal_noise_2023}, we focus on exploring the advantages and limitations of the algorithm from the classical Fisher information perspective drawing upon relevant methodological developments in QAE~\cite{tanaka_amplitude_2021,tanaka_noisy_2022}. 
We start by defining the estimation problem, reviewing the theory of RAE, and presenting the details of our hardware experiments and inference protocols in Sec.~\ref{sec:tech_background}. In Sec.~\ref{sec:res}, we provide a context for choosing {\it ibmq\_montreal} to demonstrate RAE and report the results of the one- and two-qubit RAE experiments followed by the analysis of factors impacting RAE performance. 
Finally, in Sec.~\ref{sec:concl}, we summarize our findings and discuss future research directions.

\section{Technical Background}
\label{sec:tech_background}

%Brief summary of how RAE works and reference to previous publications.

RAE is a technique that leverages low-depth quantum amplitude estimation to maximize information gain per measurement sample~\cite{wang2020bayesian}. 
This is achieved by augmenting a state preparation circuit with a series of layers called Grover iterates.
These layers enhance the sensitivity of measurement outcomes to the unknown expectation value compared to the standard sampling procedure of VQE.
In a noiseless setting, this changes the scaling of estimation runtime with precision $\epsilon$ from $\mathcal{O}(1/\epsilon^2)$ (standard quantum limit) to $\mathcal{O}(1/\epsilon)$ (Heisenberg limit), but in practice, the performance of the algorithm is curbed by noise and critically depends on how closely the noise profile of a realistic device resembles the exponential decay model. 

\subsection{$H_2$ Molecular Hamiltonians}
\label{subsec:problem_defs}
In this study, we focus on the problem of estimating the ground state energy of the $H_2$ molecule, considering it in the STO-3G minimal basis. The ground state wave function of $H_2$ in this basis is a linear combination of two Slater determinants with doubly occupied $\sigma_g$ and $\sigma_u^{*}$ orbitals due to spin and symmetry constraints. To represent it on a quantum computer, one can either directly map it on a two-level problem encoded with one qubit or derive its two-qubit counterpart starting with a standard Jordan-Wigner mapping and performing qubit tapering~\cite{Hempel2018}.

In the first case, the Hamiltonian takes the form:
\begin{align}
   \label{eq:h_1q}
   H = a^{(1)} + b^{(1)} X_0 + c^{(1)} Z_0,
\end{align}
where the parameters $a^{(1)}$, $b^{(1)}$, and $c^{(1)}$ were set to $-0.329$, $0.181$, $-0.788$, respectively. To estimate the ground state energy of the above Hamiltonian, we employ a single $Y$-rotation ansatz, optimized on a noiseless simulator: 
\begin{align}
    |\Psi(\theta)\rangle = e^{-i\frac{\theta}{2}Y_0} |0\rangle,
\end{align}
with $\theta = -6.5095$. The expectation values of Pauli operators in Eq.~\ref{eq:h_1q} are $\langle X_0 \rangle = \sin(\theta) = -0.2243$ and $\langle Z_0 \rangle = \cos(\theta) = 0.9745$. 

In the second case, we express the Hamiltonian as follows:
\begin{align}
    \label{eq:h_2q}
    H = a^{(2)} + b^{(2)} Z_0 + c^{(2)} Z_1 + d^{(2)} Z_0 Z_1 + e^{(2)} X_0 X_1 + f^{(2)} Y_0 Y_1,
\end{align}
with parameters $a^{(2)} = 0.2388$, $b^{(2)} = 0.3466$, $c^{(2)} = -0.4439$, $d^{(2)} = 0.5736$, $e^{(2)} = 0.09075$, and $f^{(2)} = 0.09075$. 
Following Ref.~\cite{Hempel2018} and utilizing the Unitary Coupled-Cluster (UCC) ansatz~\cite{romero_strategies_2018, Hempel2018}, we arrive at the following form of the ground state wave function~\footnote{It is assumed that the qubits are numbered from right to left, i.e. $|i_n, i_{n-1}, ..., i_0\rangle $.}:
\begin{align}
|\Psi(\theta)\rangle = e^{-i\frac{\theta}{2} X_1 Y_0} |01\rangle,    
\end{align}
where $\theta$ is set to its optimal value $-6.0575$. The expectation values of the Pauli strings in our two-qubit Hamiltonian (Eq.~\ref{eq:h_2q}) can be evaluated analytically as functions of $\theta$, resulting in $\langle Z_0 \rangle = -\cos(\theta) = -0.9746$, $\langle Z_1 \rangle = \cos(\theta) = 0.9746$, $\langle Z_0Z_1 \rangle = 1$, and $\langle X_0X_1 \rangle = \langle Y_0Y_1 \rangle = -\sin(\theta) = -0.2238$.

%Present the problem: H2, and the Ansatze used: 1-qubit representation and 2-qubit representation.
\subsection{RAE Ground State Energy Estimation}
To estimate the ground state energy with RAE, one needs to evaluate the expectation values of individual unitary fragments of the corresponding Hamiltonian with respect to the ground state prepared by an ansatz circuit. Throughout this work, we used the most straightforward decomposition technique, in which unitary fragments are the Pauli strings. Each string requires a separate set of RAE circuits, even within the groups of commeasurable operators~\cite{zhao_measurement_2020, min_clique_grouping, linear_grouping_algo, yen2021cartan}.

\begin{figure}
\center
\includegraphics[width=0.8\linewidth]{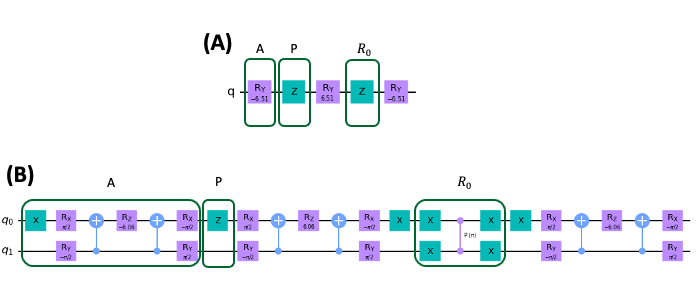}
\caption{Annotated RAE circuits with one Grover layer for estimating $Z_0$ operator for (A) one- and (B) two-qubit problems. Circuit elements (marked with $A$, $P$, and $R_0$) are defined in the main text.} 
\label{fig:mpea_comp}
\end{figure}

A RAE circuit consists of an ansatz block $A$ followed by a series of $L$ Grover layers. The Grover layers are composed of a Pauli string $P$, whose expectation value needs measurement, and a reflection with respect to the state prepared by the circuit $A$, $R_A = 2|A \rangle \langle A | - \mathbb{I}$. 
The reflection operator is often represented using $A$ and the "phase oracle" $R_0 = 2|0 \rangle \langle 0 | - \mathbb{I}$, $R_A = AR_0A^{\dagger}$, where $R_0$ acts as the identity operator on the state $| 0 \rangle$ and multiplies any orthogonal state by $-1$~\footnote{Phase oracle can be implemented as a multi-controlled $Z$ operation sandwiched between two layers of $X$ gates. Depending on the device connectivity, it can be further compiled into a set of two-qubit gates, the number of which scales at least quadratically with the number of qubits~\cite{johnson_reducing_2022}.}. 
The Grover layer sequence concludes with context selection gates chosen to rotate the final state to the eigenbasis of the Pauli string $P$. Examples of RAE circuits with one Grover layer are presented in Fig.~\ref{fig:mpea_comp}. 

More advanced methods, such as anti-commuting grouping~\cite{izmaylov_unitary_2019}, can boost RAE efficiency by reducing the number of expectation values to be evaluated. The corresponding RAE circuits will have the same structure as described above, with the Pauli operator $P$ replaced by a more complicated unitary. The benefit of grouping can, therefore, be compromised by increased circuit depth. Exploring such trade-offs is beyond the scope of this work. 

\subsection{Observable Estimation with Noise}
\label{subsec:rae_overview}
Accurate estimation of observables from noisy quantum measurements is a fundamental task in quantum algorithms. Typically, in hardware implementations of VQE, measurements are performed on the qubit register in the standard basis (i.e., the basis of Pauli $Z$ operators) to obtain the expectation values of individual Pauli strings $\Pi_i$ or groups of co-measurable Pauli terms comprising an observable of interest $O = \sum_i c_i \Pi_i$ with respect to an ansatz state. In the simplest case, where a single Pauli string is considered, the measurement outcomes are modeled as Bernoulli random variables assuming values $\pm 1$ with probabilities $p(\pm 1|\Pi) = \frac{1}{2}(1 \pm \Pi)$, where $\Pi$ represents the unknown expectation value. After accumulating $N$ samples, various approaches can be employed to estimate $\Pi$. 

In what we will refer to as the \textit{standard sampling}, the measurement outcomes are averaged to obtain an estimate of the unknown expectation $\hat{\Pi} = \frac{N_+ - N_-}{N}$, where $N_{\pm}$ is the number of $\pm 1$ among $N$ measurement samples.

%The estimator obtained through MLE is known to be unbiased. To evaluate its variance, we employ a statistical technique known as bootstrapping. In bootstrapping, $N$ measurement outcomes are resampled with repetitions from the original dataset, and MLE is applied to the corresponding posterior distributions. The variance of $\hat{\Pi}$ is then calculated from the inferred observable values obtained throughout this process.

An alternative is to adopt a \textit{Maximum Likelihood Estimation} (MLE) framework~\cite{wasserman_all_2004}, wherein the unknown parameter is estimated by maximizing the total likelihood $\mathcal{L}$ of the sample set defined as a product of the probabilities for individual samples:
\begin{align}
\label{eq:mle_1par}
\mathcal{L}(\mathbf{d}; \Pi) = \prod_{i = 1}^N p(d_i|\Pi),   
\end{align}
where $\mathbf{d} = (d_1, d_2, ..., d_N)$ and $d_i = \pm 1$.  In practice, however, it is more convenient to maximize the log-likelihood defined as $l(\mathbf{d}; \Pi) = \log \mathcal{L}(\mathbf{d}; \Pi)$, yielding an estimate for $\Pi$, $\hat{\Pi} = \operatorname{argmax}_{\Pi} l(\mathbf{d}; \Pi)$.\footnote{$\operatorname{argmax}_{y}f(x;y)$ denotes the maximum value of $f$ with respect to $y$ for fixed $x$. 
} Substituting the expression for $p(\pm 1|\Pi)$ in Eq.~\ref{eq:mle_1par} and finding the maximum of $l(\mathbf{d}; \Pi)$, it is straightforward to show that in this case, MLE is equivalent to the standard sampling. More generally, it can be applied to complicated estimation problems with multiple unknown parameters, and non-linear likelihood functions encountered in RAE, in which case $\Pi$ is replaced with a vector $\mathbf{\Pi}$, and maximization needs to be carried out in a multi-dimensional space. Provided the estimate $\hat{\mathbf{\Pi}}$ is unbiased, the covariance matrix of the estimation error $\operatorname{cov}(\hat{\mathbf{\Pi}}) = \mathbb{E}\left[ (\hat{\mathbf{\Pi}} -\mathbf{\Pi})(\hat{\mathbf{\Pi}} -\mathbf{\Pi})^T \right]$ \footnote{$\mathbb{E}(A)$
denotes the element-wise expectation value matrix of the matrix $A$, i.e. $\mathbb{E}(A)_{ij}=\mathbb{E}(A_{ij})$.} satisfies the Cram\'{e}r-Rao (CR) bound
\begin{align}
    \label{eq:cr_bound}
    \operatorname{cov}(\hat{\boldsymbol{\Pi}}) \ge \mathcal{I}^{-1}(\boldsymbol{\Pi}),
\end{align}
where $\mathcal{I}(\boldsymbol{\Pi})$ is the Fisher information matrix with the matrix elements
\begin{align*}
    [\mathcal{I}(\boldsymbol{\Pi})]_{i, j}=\mathbb{E}\left[\left(\frac{\partial}{\partial \Pi_i} \ln \mathcal{L}(\mathbf{d} ; \boldsymbol{\Pi})\right)\left(\frac{\partial}{\partial \Pi_j} \ln \mathcal{L}(\mathbf{d} ; \boldsymbol{\Pi})\right)\right].
\end{align*}
%Fisher information is known to be additive in the number of samples. 
If the CR bound is tight, the mean squared estimation error $\mathbb{E}\left[(\hat{\Pi} - \Pi)^2\right]$ is inversely proportional to Fisher information. Therefore, by maximizing information gain per sample, one can improve the efficiency of the estimation technique. In the context of RAE, this is achieved by modifying the likelihood function in a way that makes it more sensitive to the parameter of interest at the expense of increasing circuit depth. Under realistic experiment conditions, quantum circuits are subject to decoherence, limiting the maximum number of Grover iterates and the sampling boost offered by RAE. To describe noise effects on the estimation, we followed the standard approach adopted by Wang et al.~\cite{wang2020bayesian} and assumed the global depolarizing noise under which an arbitrary n-qubit state $\rho$ transforms into
\begin{align}
    \mathcal{E}_d(\rho) = p \rho + \frac{(1-p)}{2^n}\mathbb{I},
\end{align}
where $p \in [0, 1]$ is the fidelity. If the action of the noise channel $\mathcal{E}_d$ is interleaved with applying Grover iterates $U = R_AP$ (i.e. $p$ is the Grover layer fidelity), an $L$-layer RAE circuit prepares a state
\begin{align}
\rho_L = p^L U^L |A\rangle \langle A| (U^{\dagger})^L + \frac{(1-p^L)}{2^n}\mathbb{I}.
\end{align}
The first term in the above equation contains information about the observable of interest. Its amplitude decreases exponentially with $L$, following $p^L$. %For consistency with the previous work~\cite{wang2020bayesian,johnson_reducing_2022,katabarwa_reducing_2021,dalal_noise_2023} we introduce the effective noise parameter $\lambda$, related to the Grover layer fidelity $p$ as $\lambda = -\ln(p)$. 
Although depolarizing noise may not accurately describe realistic quantum devices~\cite{dalal_noise_2023}, it allows one to derive an analytic expression for the distribution of measurement outcomes, referred to as bitstring parities, from an $L$ layer RAE circuit~\cite{wang2020bayesian}:
\begin{align}
    \label{eq:cheb_likelihood_gen}
       \mathcal{P}_L(d|\Pi, \lambda) &= \frac{1}{2} \Big(1 + (-1)^{d}Tr[\rho_LP] \Big) \\\label{eq:cheb_likelihood}
       &= \frac{1}{2} \Big(1 + (-1)^{d} e^{-\lambda(L + 1/2)} T_{2L + 1}(\Pi)\Big),
\end{align}
In Eq.~\ref{eq:cheb_likelihood}, $\Pi$ is the expectation value of $P$ with respect to the ansatz state, $\Pi = \langle A | P | A \rangle$,  $d$ is the bitstring parity ($d \in \{0, 1\}$), and $T_{2L + 1}(\Pi) = \cos\{(2L + 1) \operatorname{acos} \Pi \}$ is the Chebyshev polynomial of the first kind. The effective noise parameter $\lambda$, is related to the Grover layer fidelity $p$, with $\lambda = -\ln(p)$. The exponent of the $\lambda$-dependent damping factor implies that the quantum state experiences twice as much decoherence from a single application of $R_A$ as it does from the ansatz circuit $A$.  

The likelihood function of RAE is a significant improvement over the one considered at the beginning of this section in several aspects. First, it explicitly incorporates noise via a single nuisance parameter $\lambda$. Second, it introduces an additional degree of freedom $L$ to maximize Fisher information and reduce the MLE error. Non-linearity of $\mathcal{P}_L(d|\Pi, \lambda)$ gives rise to ambiguity in estimation since multiple values of $\Pi$ can correspond to the same likelihood value. To ensure the uniqueness of the $\Pi$ estimate, we combined MLE with a circuit fusion technique developed for low-depth QAE~\cite{suzuki_amplitude_2020,tanaka_amplitude_2021} in which the total likelihood is computed as a product of the likelihood functions corresponding to varying Grover depth, $L$.  The ordered set of $L$ values, referred to as a "layer schedule", can be treated as a hyperparameter of the algorithm along with the number of measurement samples for each $L$~\cite{tanaka_amplitude_2021,brown_quantum_2020,suzuki_amplitude_2020,callison_improved_2023,giurgica-tiron_low-depth_2022}. The two commonly used layer schedules are "linearly incremental sequence" (LIS)
\begin{align}
    L_i = i,
\end{align}
and "exponentially incremental sequence" (EIS)
\begin{align}
   L_i = \lfloor 2^{i - 1} \rfloor,
\end{align}
where $i = 0, 1, 2, ...$. In a noiseless setting, if the number of samples is constant for different $L_i$, EIS results in the estimation runtime asymptotically approaching the Heisenberg limit, i.e., scaling as $\mathcal{O}(1/\epsilon)$. By contrast, LIS corresponds to an intermediate regime between standard sampling and Heisenberg limit, $\mathcal{O}(1/\epsilon^{4/3})$. For a general polynomial schedule of degree $d$, i.e. $L_i = i^d$, the runtime scales as $\mathcal{O}(1/\epsilon^{(2d + 2)/(2d + 1)})$. As pointed out by Brown et al.~\cite{brown_quantum_2020}, EIS yields a rapid increase in circuit depth and, as a result, tends to give a larger asymptotic error compared to LIS in the presence of noise. Away from the asymptotic regime EIS, however, can be superior to LIS. For this reason, we checked that LIS is equivalent to EIS in terms of the minimum estimation error for one of our experiments (Appendix~\ref{app:schedules}) and used the former throughout this work.

For any layer schedule $\mathbf{L}$ under the assumption of $N_s$ samples per circuit, the total likelihood takes the form:
\begin{align}
\label{eq:rae_likelihood}
    \mathcal{L}(\mathbf{d}; \Pi, \lambda) = \prod_{L \in \mathbf{L}} \mathcal{P}_L(0|\Pi, \lambda)^{e_L} \mathcal{P}_L(1|\Pi, \lambda)^{N_s - e_L}. 
\end{align} 
Here, $e_L$ denotes the number of even bit-strings observed for circuits with $L$ layers. Using additivity of Fisher information with respect to the number of samples, one can derive the elements of the $\mathcal{I}(\Pi, \lambda)$ matrix corresponding to the likelihood function $\mathcal{L}(\mathbf{d}; \Pi, \lambda)$:
\begin{align}
\label{eq:fisher_info11}
   \mathcal{I}(\Pi, \lambda)_{11} &= \sum_{L\in\mathbf{L}} \frac{N_s\left(2L + 1\right)^{2} \sin^{2}{\left(\left(2L + 1\right) \operatorname{acos}{\left(\Pi \right)} \right)}}{\left(1 - \Pi^{2}\right) \left(e^{\lambda \left(2L + 1\right)} - \cos^2{\left(\left(2 L + 1\right) \operatorname{acos}{\left(\Pi \right)} \right)}\right)}\\
   \label{eq:fisher_info12}
    \mathcal{I}(\Pi, \lambda)_{12} &= \sum_{L\in\mathbf{L}} \frac{N_s\left(L + 1/2\right)^2 \sin{\left(2\left(2 L + 1\right) \operatorname{acos}{\left(\Pi \right)} \right)}}{\sqrt{1 - \Pi^{2}} \left(- e^{\lambda \left(2L + 1\right)} + \cos^{2}{\left(\left(2 L + 1\right) \operatorname{acos}{\left(\Pi \right)} \right)}\right)}\\
    \label{eq:fisher_info22}
    \mathcal{I}(\Pi, \lambda)_{22} &= \sum_{L\in\mathbf{L}} \frac{N_s\left(L + 1/2\right)^2 \cos^{2}{\left(\left(2 L + 1\right) \operatorname{acos}{\left(\Pi \right)} \right)}}{e^{\lambda \left(2L + 1\right)} - \cos^{2}{\left(\left(2 L + 1\right) \operatorname{acos}{\left(\Pi \right)} \right)}}
\end{align}
Note that $\mathcal{I}(\Pi, \lambda)$ is singular if $\mathbf{L} = \{ 0 \}$ for any $\lambda$ and $\Pi$ since the unknown parameters are not uniquely identifiable from MLE\footnote{This also follows from Eq.~\ref{eq:cheb_likelihood}. Indeed, for $L = 0$, it can be shown that applying MLE results in an underdetermined system of equations.}. Adding extra measurements from the circuits with $L \ne 0$ removes the degeneracy, making $\mathcal{I}(\Pi, \lambda)$ invertible.  

Assuming prior knowledge of $\lambda$ and $\Pi$, one can use Eqs.~\ref{eq:fisher_info11}-\ref{eq:fisher_info22} to compute the theoretical lower bound for the $MSE(\hat{\Pi})_{RAE}$\footnote{In the following we use notation $MSE(\hat{\Pi})$ for the mean squared error of $\hat{\Pi}$, $\mathbb{E}\left[(\hat{\Pi} - \Pi)^2\right]$, with the subscript referring to the estimation method.} according to the CR inequality. Similarly, it is straightforward to evaluate the MSE for the direct sampling estimator, establishing a reference for the RAE performance. 
In the presence of depolarizing noise, the ansatz circuit prepares the state
\begin{align}
    \rho_0 = e^{-\lambda/2} |A \rangle \langle A | + \frac{1-e^{-\lambda/2}}{4}\mathbb{I}.
\end{align}
If one takes $N_s$ measurements of the Pauli string $P$ in state $\rho_0$ and averages the outcomes to estimate  $\Pi$, the MSE of such estimator is
\begin{align}
    MSE(\hat{\Pi})_d = (1-e^{-\lambda/2})^2\Pi^2 + \frac{1 - e^{-\lambda}\Pi^2}{N_s}
    \label{eq:vqe_perf_model}
\end{align}
The CR bound combined with Eq.~\ref{eq:vqe_perf_model} yeilds a criterion to detect RAE sampling boost from experimental data. Specifically, we consider RAE to be successful in accelerating sampling on a real device whenever the empirical root mean squared error of the expectation of $P$, $RMSE(\hat{\Pi})$, is smaller than $\sqrt{MSE(\hat{\Pi})_d}$ for the {\it same number of ansatz queries}, i.e., the inequality 
\begin{align}
    \sqrt{MSE(\hat{\Pi})_{RAE}} \le RMSE(\hat{\Pi}) < \sqrt{MSE(\hat{\Pi})_d}
    \label{eq:rae_adv_criterion}
\end{align}
is satisfied with statistical accuracy. In practice, statistical error of the empirical $RMSE(\hat{\Pi})$ can either be estimated from a series of independent experiments or by performing bootstrapping on a single set of experimental results.

When applying RAE to measuring the expectation of a multi-term operator $O = \sum_i c_i \Pi_i$, it is important to consider an optimal shot allocation for measuring the expectations of Pauli terms it is comprised of. Although a solution to this problem is well known for direct sampling, the equivalent assignment of the number of circuit repetitions per Pauli $N_{s}^{(i)}$ is generally not available for RAE since $RMSE(\hat{\Pi}_i)$ is not known analytically as a function $N_{s}^{(i)}$. Although it is possible to establish an approximate runtime model for the single Pauli term estimation~\cite{johnson_reducing_2022} and use it for measurement analysis, we chose to apply the uniform shot allocation for both RAE and direct sampling when calculating ground state energy, i.e., $N_{s}^{(i)} = N_s$. A more detailed analysis of the measurement allocation is left for future work. 

%It is important to note that MLE based on Eq.~\ref{eq:rae_likelihood} helps to avoid aliasing (i.e., multiple solutions) by incorporating samples from RAE circuits with $L = 0$\footnote{Unknown expectation value and the noise parameter cannot be estimated unambiguously from such samples alone. For example, MLE produces an infinite number of $(\Pi, \lambda)$ pairs, such that $\lambda = 2\ln(\Pi)$ for $d = 0$ and $L = 0$, as can be seen from Eq.~\ref{eq:cheb_likelihood}. When combined with measurement results for $L \ne 0$, the posterior will have a unique maximum, provided the number of shots is large enough.}.
%\todo{add a discussion of shot allocation}
%We applied bootstrapping to compute the standard deviation of the estimator in Eq.~\ref{eq:mle}~\cite{wasserman_all_2004}. To ensure the samples from circuits with $L = 0 - L_{max}$ are always included when computing $p(\Pi, \lambda)$, the resampling was performed within subsets of bitstring parities for each $L$ individually. The distributions were represented on a grid with 100 and 1000 points for $\lambda$ and $\Pi$ axes, respectively. After conducting a series of $B$ bootstrap simulations, the expectation values and corresponding standard deviations of $\Pi$ and $\lambda$ are computed using standard expressions~\cite{wasserman_all_2004}.

\subsection{Details of the experiments and post-processing}
\label{subsec:exp_details}

To evaluate RAE performance, we conducted two series of hardware experiments. In the first one, meant to probe the validity of the depolarizing noise model, we estimated the likelihood of getting even bitstings when measuring  $X_0X_1$,  $\mathcal{P}_L(0|\Pi, \lambda)$ for $L = 1 - 5$ and 10 uniformly spaced values $\Pi \in [0, 1]$. 
The RAE circuits were implemented on \textit{ibmq\_hanoi} (QV\footnote{QV = Quantum Volume~\cite{cross_validating_2019,blume-kohout_volumetric_2020}} 64), \textit{ibmq\_sydney} (QV 32), \textit{ibmq\_toronto} (QV 32), and \textit{ibmq\_montreal} (QV 128) quantum devices available on IBM quantum cloud as of 2021. 8192 shots were requested in each experiment. A readout correction was applied using the same number of shots to evaluate the measurement filters~\cite{bravyi_mitigating_2021} for the appropriate qubit subsets. To reconstruct $\mathcal{P}_L(0|\Pi, \lambda)$ from our experiments we estimated the expectation values of $X_0X_1$ with respect to $\rho_L$ and applied Eq.~\ref{eq:cheb_likelihood_gen}. The error bounds were computed from empirical standard deviations of the $X_0X_1$ expectation values. Chebyshev likelihood functions (Eq.~\ref{eq:cheb_likelihood}) were fitted to experimental data to extract effective noise parameters $\lambda$ using \texttt{scipy.optimize.curve\_fit}~\cite{2020SciPy-NMeth}. The standard error of $\lambda$, $\delta\lambda$, was computed as a square root of its variance estimated using a linear approximation to the model function near the optimum. In the second series of experiments, we estimated the ground state energies of the one- and two-qubit Hamiltonians (Eq.~\ref{eq:h_1q} and Eq.~\ref{eq:h_2q}) on \textit{ibmq\_montreal} with RAE. Experiments involved submitting RAE circuits created individually for each term in corresponding Hamiltonians and terminated with the appropriate context selection gates. The maximum number of Grover iterates was set to 10 and 8 for one- and two-qubit circuits, respectively, with 8192 samples per circuit. As described in Sec.~\ref{subsec:rae_overview}, the expectation values of the Pauli terms were extracted from experimental data via MLE. To determine the error bounds of the estimates, we applied \textit{bootstrapping}~\cite{wasserman_all_2004,efron_introduction_1994} following Refs.~\cite{katabarwa_reducing_2021,dalal_noise_2023}. Specifically, given a layer schedule $\mathbf{L}$, we generated a list of maximum likelihood estimates for each Pauli term $P$ by resampling 8192 bitstring parities with repetitions from each $L \in \mathbf{L}$ and using them to compute and maximize the log-likelihood $l(\mathbf{d}; \Pi, \lambda)$. Given the list of $M$ estimates $\{\hat{\Pi}_i\}_1^M$ we computed $MSE(\hat{\Pi})$ as 
\begin{align}
    MSE(\hat{\Pi}) = \frac{1}{M} \sum_i (\Pi_i - \Pi)^2
\end{align}
The empirical variance of $MSE(\hat{\Pi})$ was taken as a measure of its statistical error and was evaluated as 
\begin{align}
    \mathbb{Var}(MSE(\hat{\Pi})) = \frac{1}{M} \sum_i \big\{ (\Pi_i - \Pi)^2 - MSE(\hat{\Pi}) \big \}^2
\end{align}
Standard error propagation formula was applied to $\mathbb{Var}(MSE(\hat{\Pi}))$ to obtain statistical error of $RMSE(\hat{\Pi}) = \sqrt{MSE(\hat{\Pi})}$:
\begin{align}
    \sigma = \frac{\sqrt{\mathbb{Var}(MSE(\hat{\Pi}))}}{2\cdot RMSE(\hat{\Pi})}
\end{align}
When post-processing the results of RAE experiments, we chose $M = 15000$ and $M = 10000$ for one-qubit and two-qubit circuits, respectively.

In all cases, the circuits were submitted via Qiskit~\cite{javadi-abhari_quantum_2024} with the optimization level set to 0, which resulted in the transpiler expressing the circuits in the native gate set and mapping them to qubits 0 and 1.

%To aid the analysis of the two-qubit experiments, we grouped the measurement outcomes into three categories based on the type of Pauli terms and how much device noise was expected to accrue during the execution of the corresponding circuit. The first group included one-qubit terms, $Z_0$ and $Z_1$; the second group was comprised of the two-qubit terms, $X_0X_1$ and $Y_0Y_1$, excluding $Z_0Z_1$ term, which served as a probe for device noise, as its expectation is independent of the ansatz parameter and is equal to -1. 

%Additionally, to get insights into the QPU stability and noise levels, we generated experimental likelihood functions for the $X_0X_1$ term of the two-qubit Hamiltonian~\ref{eq:h_2q} and the number of Grover iterates $L$ ranging from 1 to 5. The ansatz parameter $\theta$ was tuned from $0$ to $-\frac{\pi}{2}$ to ensure the exact expectation value $\langle X_0X_1 \rangle$ varies from 0 to 1. For each value of $\langle X_0X_1 \rangle$ the likelihood of measuring an even bitstring (corresponding to $d = 0$ in Eq.~\ref{eq:cheb_likelihood}) was estimated based on 8192 shots\footnote{The error bounds of the estimates were calculated as sample standard deviations}. Chebyshev likelihood functions were then fitted to experimental data to extract effective noise parameters $\lambda$. The estimates of $\lambda$ we used to determine appropriate inference intervals in MLE as well as to assess RAE performance as demonstrated in the following Section. 

\section{Results}
\label{sec:res}

The goal of our experiments was to probe several aspects of RAE, including (a) the impact of realistic hardware noise on the convergence of expectation values, (b) sampling rate increase compared to the standard approach, and (c) the error mitigation properties reported by Dalal et al.~\cite{dalal_noise_2023}. Below, we commented on the aforementioned aspects of RAE in the context of one- and two-qubit ground-state energy estimation problems considered in this work. As a prerequisite, we investigated the relevant properties of the quantum devices considered in this work and extracted the effective noise parameter $\lambda$ from RAE data for \textit{ibmq\_montreal} in Sec.~\ref{subsec:likelihood} and Sec.~\ref{subsec:eff_lambda}. The values of $\lambda$ were used as inputs to the direct sampling and RAE performance models discussed in Sec.~\ref{subsec:rae_application}.

%This informed our inference protocol for later experiments, allowing to define suitable priors for $\lambda$. In the following, we studied the convergence of Pauli term expectations to compute the ground state energies of one- and two-qubit model problems as functions of RAE circuit depth. To compare the device performance, we determined the maximum number of Grover layers beyond which it is impossible to extract information about the observable of interest. To probe the stability of the results with respect to the fluctuation of device parameters~\cite{Dasgupta2021} some experiments were repeated several times. Finally, the sampling rate was evaluated assuming the total number of ansatz queries $N_q$ as a proxy for the runtime. Using standard sampling the number of state preparations required to reach a given precision scales as $\mathcal{O}(1/\sqrt{N_q})$, whereas in RAE we expect to find  $\mathcal{O}(1/N_q^{\alpha})$ with $\alpha > 1/2$. In particular, for the LIS $\alpha$ is predicted to be $3/4$ closely approaching the Heisenberg limit corresponding to $\alpha = 1$.

\subsection{Noisy likelihood functions and RAE performance}
\label{subsec:likelihood}
\begin{figure}[!h]
\center
\includegraphics[width=17.5cm]
{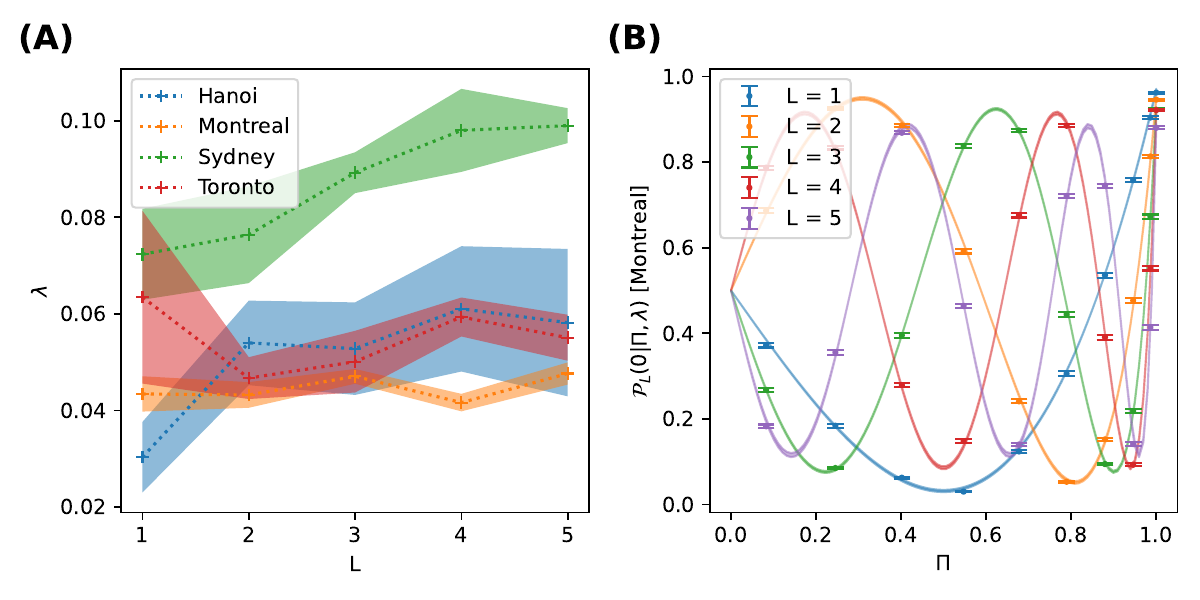}
\caption{(Panel (A): effective noise parameters $\lambda$ as functions of the RAE circuit depth (the number of Grover iterates, $L$) on IBM quantum devices. Shaded areas correspond to $\lambda \pm \delta \lambda$ values obtained via non-linear fit. Panel (B): measured likelihood function (points with error bars) fitted to $\mathcal{P}_L(0|\Pi, \lambda)$ (solid lines) for $L = 1 - 5$ on \textit{ibmq\_montreal}. The numerical values of $\lambda$ and their errors are presented in Tab.~\ref{tab:lambda_eff_premium}. Line width demonstrates the variation of the likelihood function with the noise parameter $\in [\lambda - \delta \lambda, \lambda + \delta \lambda]$} 
\label{fig:lambda_values}
\end{figure}

A preliminary assessment of the validity of depolarizing noise as a model for decoherence effects in our experiments can be made based on how well Eq.~\ref{eq:cheb_likelihood} describes empirical likelihood functions. We found that it captures their overall shapes with some discrepancies, the magnitude of which differs between the devices and can be quantified by the standard errors of the $\lambda$ values extracted from non-linear fits. As one can see in Fig.~\ref{fig:lambda_values} (Panel A), the depolarizing noise model is the least accurate for \textit{ibmq\_hanoi}, and its accuracy does not strongly correlate with the QV.  Furthermore, the values of $\lambda$ vary with the number of Grover iterates.

This behavior has not been previously observed in the context of RAE and is likely due to the device parameter instability reported by other authors in a more general benchmarking setting~\cite{Dasgupta2021}. Since the RAE inference protocol assumes a noise parameter independent of $L$, we expect a noticeable impact on the algorithm performance due to the variation of $\lambda$. In principle, this restriction can be removed with a more flexible multi-parameter likelihood function~\cite{tanaka_noisy_2022} at the expense of more complicated post-processing and a different experimental setup relying on additional assumptions about the noise model. 

Comparing the degree of $\lambda$ variation between the devices considered in this work, we note that  \textit{ibmq\_montreal} is the most stable ($\lambda$ changes by at most 14\%), while the performance of the other three QPUs is roughly similar (with $\lambda$ varying by about 50\% in some cases). Another favorable feature of \textit{ibmq\_montreal} is the close agreement between its likelihood function and Eq.~\ref{eq:cheb_likelihood} as can be seen in Fig.~\ref{fig:lambda_values} (Panel B) suggesting the validity of the depolarizing noise model for this device.

For these reasons, we chose  \textit{ibmq\_montreal} to demonstrate RAE performance. In the following sections, we showed that the criterion in Eq.~\ref{eq:rae_adv_criterion} was satisfied in most of the RAE experiments on that backend, indicating a sampling acceleration compared to the direct method commonly used in VQE. Furthermore, RAE demonstrated noise mitigation properties in some cases, allowing RMSE reduction for $Z_0$ and $Z_1$ Pauli terms in the two-qubit Hamiltonian by about one order of magnitude compared to the asymptotic direct sampling limit. The combined effect of sampling boost and noise mitigation resulted in ground state energy estimates within chemical accuracy for both one- and two-qubit problems. This contrasts with the results from direct sampling for the two-qubit problem and marks a significant improvement in accuracy for the one-qubit problem.

\subsection{Effective noise parameters and direct sampling performance}
\label{subsec:eff_lambda}
\begin{figure}
\center
\includegraphics{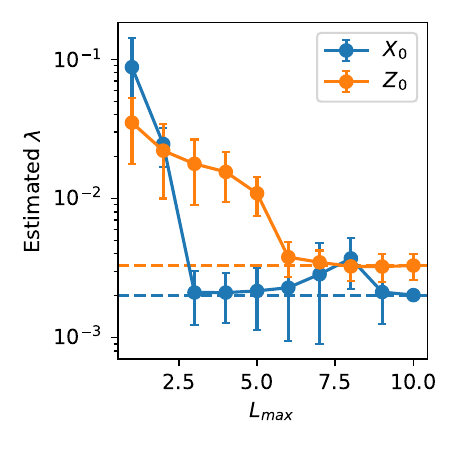}
\caption{Convergence of the $\lambda$ values estimated via MLE for the Pauli terms in the one-qubit Hamiltonian (Eq.~\ref{eq:h_1q}). The error bars were calculated as standard deviations of the bootstrapped $\hat{\lambda}$ values as explained in Sec.~\ref{subsec:exp_details}.  $L_{max}$ refers to the maximum Grover depth used to estimate $\lambda$. Dashed lines indicate the converged values of the noise parameter.} 
\label{fig:mle_lambda_1q_montreal}
\end{figure}

\begin{figure}
\center
\includegraphics{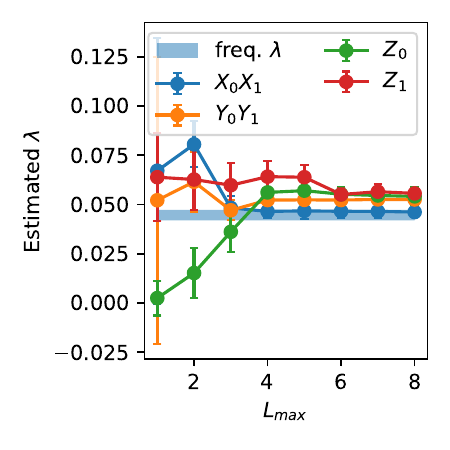}
\caption{Convergence of the $\lambda$ values estimated via MLE for the Pauli terms in the two-qubit Hamiltonian (Eq.~\ref{eq:h_2q}). $L_{max}$ refers to the maximum Grover depth used in the estimation. The boundary of the filled area corresponds to $\lambda$ $\pm \delta\lambda$, i.e., the range of noise parameter values determined by fitting the Chebyshev likelihood functions of $X_0X_1$ to the experimental estimates for $L_{max} = 1 - 5$.} 
\label{fig:mle_lambda_2q_montreal}
\end{figure}

To model direct sampling and RAE performance based on the theoretical CR bound, we extracted the effective noise parameters $\lambda$ from RAE results via MLE (Fig.~\ref{fig:mle_lambda_1q_montreal} and Fig.~\ref{fig:mle_lambda_2q_montreal}). For both one- and two-qubit cases, $\lambda$ estimates converge with the Grover depth, $L_{max}$ reaching asymptotic values that are close but not identical for different Pauli terms. This can be explained by the differences in RAE circuit transpilation. Further, the noise levels for the one-qubit circuits are about one order of magnitude lower than for the two-qubit case, in line with average one- and two-qubit gate fidelities.

MLE provides an independent way to validate the value of $\lambda$ obtained as described in the previous section for $X_0X_1$. In Fig.~\ref{fig:mle_lambda_2q_montreal}, one can see that the two methods provide consistent results within the error bounds.

In the following, we used MLE-derived $\lambda$ values to model the performance of the direct sampling with the number of ansatz queries as prescribed by Eq.~\ref{eq:vqe_perf_model}. Similarly, $\lambda$ enters the CR bound for the RAE MSE. It is, therefore, crucial to ensure that the estimates of $\lambda$ accurately represent the device performance. It can be shown that although RAE results for $L_{max} = 0$ are insufficient for simultaneous observable and noise estimation, using them as an input for MLE with $\lambda = 0$ is equivalent to the direct observable estimation. We relied on this property to derive reference expectation values for the Pauli terms in the corresponding Hamiltonians and perform additional consistency checks for the noise parameters. In Tab.~\ref{tab:vqe_perf_validation}, we compared RMSE of the Pauli expectations extracted from RAE experiments with $L = 0$ with the predictions of Eq.~\ref{eq:vqe_perf_model} assuming $\lambda$ values from MLE.
\begin{table}
\center
\begin{tabular}{lcc}
\hline
\hline
$\Pi$ &  $\epsilon$  &  $\epsilon$ (analytic)  \\
\hline
\multicolumn{3}{c}{\textit{One-qubit Hamiltonian}}\\
\hline
$\langle Z_0 \rangle$ & 0.0045(0)\footnote{statistical error is less than $10^{-4}$}\footnote{shorthand notation is used for the floating point numbers with finite statistical precision replacing X.XXXX $\pm$ 0.00YY with X.XXXX(YY)} & 0.0030 \\
$\langle X_0 \rangle$ &  0.0510(105) & 0.0108 \\
\hline
\multicolumn{3}{c}{\textit{Two-qubit Hamiltonian}}\\
\hline
$\langle Z_0 \rangle$ &  0.0083(28) & 0.0262 \\
$\langle Z_1 \rangle$ &  0.0246(3) & 0.0270 \\
$\langle Y_0Y_1 \rangle$ &  0.0119(82) & 0.0122 \\
$\langle X_0X_1 \rangle$ &  0.0200(99) & 0.0119 \\
\hline
\end{tabular}
\caption{Root mean square errors $\epsilon$ of the Pauli expectations  computed from experimental data collected in one- and two-qubit RAE experiments and estimated analytically (based on (Eq.~\ref{eq:vqe_perf_model}), i.e. $\epsilon = \sqrt{MSE(\hat{\Pi})_d}$) assuming effective noise parameters from MLE.}\label{tab:vqe_perf_validation}
\end{table}

As can be seen from the table, there exist discrepancies between experimental $\epsilon$ and its analytical estimate from Eq.~\ref{eq:vqe_perf_model} that cannot be attributed to statistical errors (for example, for $Z$ terms). In many cases where the two differ, however, the direct sampling model yields smaller RMSEs than the experiment, except for $Z_0$ and $Z_1$ terms in the two-qubit Hamiltonian. Effectively, this means that MLE underestimates the noise parameter for direct sampling, lowering the upper boundary of the RAE advantage window, as follows from the criterion in Eq.~\ref{eq:rae_adv_criterion}. For this reason, we used Eq.~\ref{eq:vqe_perf_model} with $\lambda$ from MLE when assessing the RAE sampling boost despite the associated discrepancies with a caveat that it may not be reliably detected in the experiments where we measured $Z_0$ and $Z_1$ expectation values.

\subsection{RAE application to ground state energy estimation}
\label{subsec:rae_application}

Having established the performance models for RAE and direct sampling, we analyzed our experiments from the perspective of the sampling advantage criterion Eq.~\ref{eq:rae_adv_criterion}.

\begin{figure}
\center
\includegraphics[width=17.5cm]{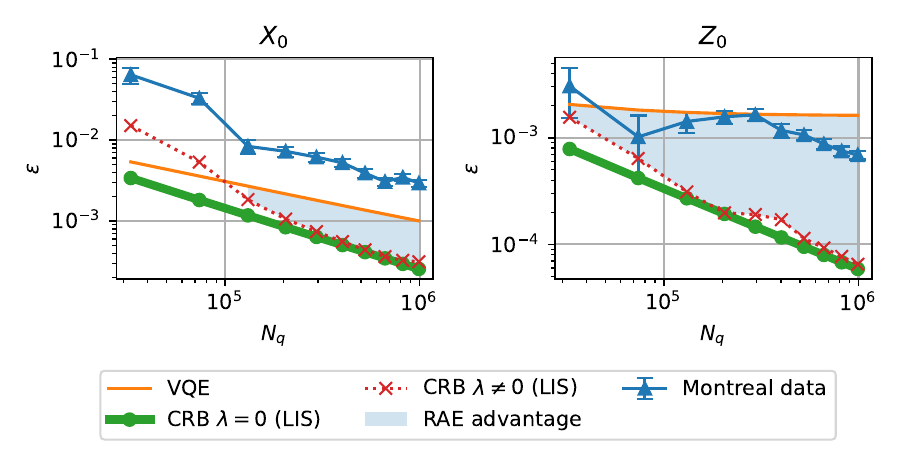}
\caption{RMSE of the estimated Pauli terms in the one-qubit Hamiltonian (Eq.~\ref{eq:h_1q}). $N_{q}$ refers to the total number of ansatz queries across all RAE circuits to obtain a Pauli term expectation. RMSE of the direct sampling estimator (marked VQE in the figure) was modeled based on the effective noise parameter extracted from RAE data. CRB refers to the lower estimate of RMSE based on the Cramer-Rao bound (Eq.~\ref{eq:cr_bound}).} 
\label{fig:rae_pauli_1q_montreal}
\end{figure}

Turning to the one-qubit case presented in Fig.~\ref{fig:rae_pauli_1q_montreal}, we observed qualitatively different behavior for $X_0$ and $Z_0$. RMSE for the former is higher than the direct sampling values, while the latter benefits from RAE sampling acceleration. Partially, this might be due to low bias for $X_0$, giving rise to a relatively small RMSE dominated by the variance. As a result, RMSE reduction gives rise to a very narrow advantage window for $X_0$. 
The noisy CR bound is far from saturation in both cases and closely agrees with the noiseless one, showing no distinct plateau~\cite{tanaka_amplitude_2021}.  Based on the form of the likelihood function, one could expect $\epsilon$ to plateau at approximately $1/\lambda \approx 300 - 500$. Accumulation of coherent error at such high Grover depth would likely compromise RAE performance long before the limit is reached. The signs of this behavior can be seen in a non-monotonous decrease of the RAE estimation error for  $Z_0$. 

Similar patterns persist for the two-qubit experiments where RAE outperforms direct sampling for the terms with high bias ($Z_0$ and $Z_1$) even at $L_{max} = 1$, while a larger Grover depth is required to observe advantage for the low-bias terms ($X_0X_1$ and $Y_0Y_1$). Although the sampling boost for $Z_0$ cannot be established reliably due to imperfections in the direct sampling performance model, one can still expect to observe the error reduction compared to direct sampling if $\epsilon = 10^{-3}$ is taken as a conservative estimate of the direct sampling error (Tab.~\ref{tab:vqe_perf_validation}). 

\begin{figure}
\center
\includegraphics[width=17.5cm]{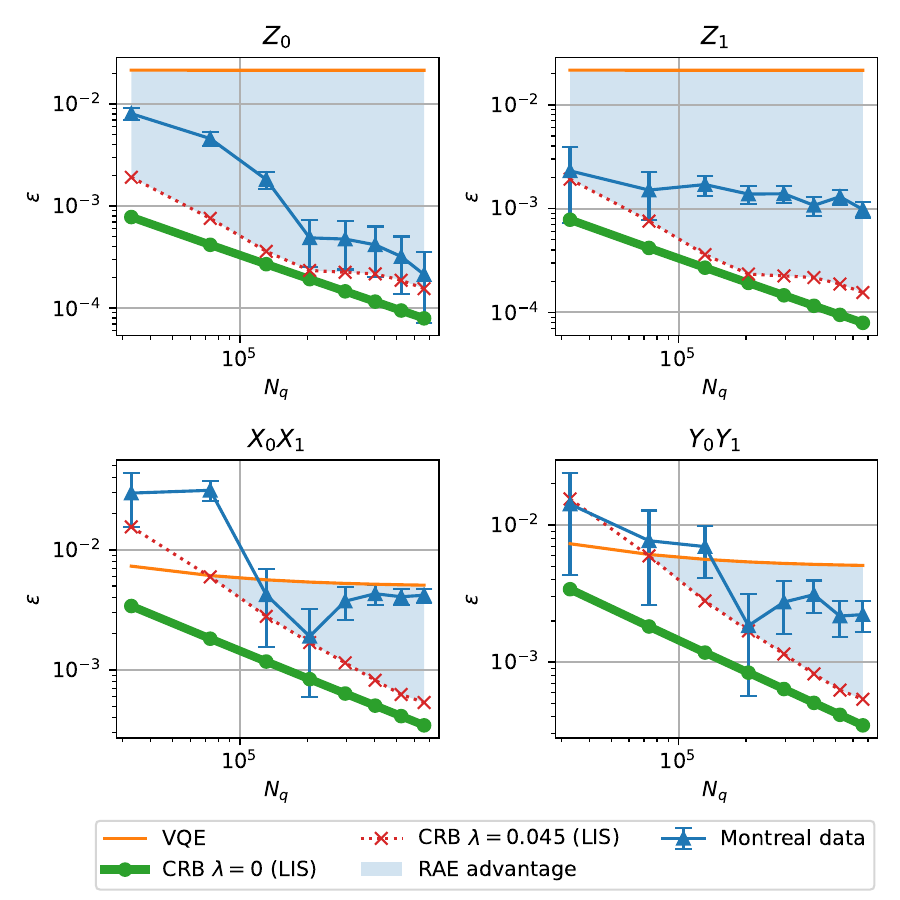}
\caption{RMSE of the estimated Pauli terms in the two-qubit Hamiltonian (Eq.~\ref{eq:h_2q}). $N_{q}$ refers to the total number of ansatz queries across all RAE circuits to obtain a Pauli term expectation. RMSE of the direct sampling estimator (marked VQE in the figure) was modeled based on the effective noise parameter from the fitted Chebyshev likelihood functions. CRB refers to the lower estimate of RMSE based on the Cramer-Rao bound (Eq.~\ref{eq:cr_bound})} 
\label{fig:rae_pauli_2q_montreal}
\end{figure}

Comparing one- and two-qubit experiments, we observed that the noise levels did not correlate with the RAE efficiency as long as the underlying decoherence mechanism was consistent with the depolarizing noise assumption. In that case, RAE can effectively learn to eliminate noise as a part of the estimation process, as suggested by Ref.~\cite{dalal_noise_2023} and corroborated by our results. In the experiments for $X_0X_1$ and $Y_0Y_1$, the minimum estimation error was achieved at the intermediate Grover depth, $L_{max} = 4$. This coincides with the value at which parameter instability developed in preliminary experiments on \textit{ibmq\_montreal} (Fig.~\ref{fig:lambda_values}), providing indirect evidence of RAE performance being sensitive to the fluctuation of $\lambda$. 

\begin{figure}
\center
\includegraphics[width=0.95\linewidth]{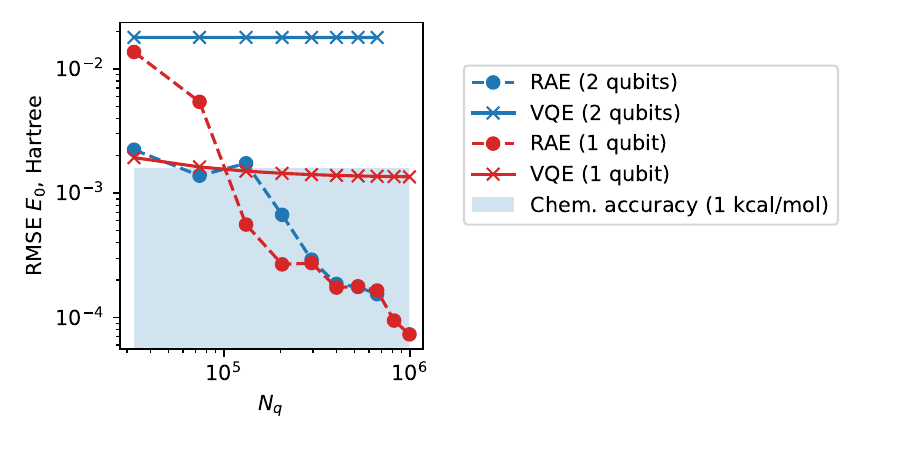}
\caption{Ground state energy RMSE for the one- the two-qubit Hamiltonians (Eq.~\ref{eq:h_1q} and Eq.~\ref{eq:h_2q}). $N_{q}$ refers to the total number of ansatz queries across all RAE circuits to obtain a Pauli term expectation assuming uniform measurement allocation. RMSE of the direct sampling estimator (marked VQE in the figure) was modeled based on the effective noise parameter (s) from the MLE and the fitted Chebyshev likelihood functions for the one- and two-qubit problems, respectively.} 
\label{fig:energy_rmse}
\end{figure}

To estimate the ground state energy with RAE, the expectation values of the individual Pauli terms were combined with appropriate weights. The variance of the estimator was computed using the standard expression for the linear combination of random variables. Energy bias and variance were then used to calculate RMSE presented in Fig.~\ref{fig:energy_rmse}. A similar calculation was performed for the direct sampling using analytic expressions for the bias and variance under the assumption of depolarizing noise with $\lambda$ parameters determined via MLE. In both cases, we assumed the uniform allocation of ansatz queries $N_q$ across Pauli terms in corresponding one- and two-qubit Hamiltonians. 

For the one-qubit Hamiltonian, despite the relatively large estimation error for the $X_0$, the RMSE ground state energy falls below chemical accuracy for the one-qubit Hamiltonian when $L_{max} > 2$ since the term enters with a smaller coefficient compared to $Z_0$. Although direct sampling reaches the chemical accuracy threshold sooner than RAE, the final RMSE for the latter is much lower, below 0.1 mH (Fig.~\ref{fig:energy_rmse}).

The ground state energy RMSE for the two-qubit Hamiltonian exhibited a similar convergence pattern, reaching chemical accuracy at $L_{max} = 2$. In this case, RAE was superior to direct sampling across the range of $N_q$ and improved the RMSE by about two orders of magnitude, as shown in Fig.~\ref{fig:energy_rmse}. Increasing the number of ansatz queries does not result in a noticeable improvement in the energy error for the direct sampling due to the large bias. In contrast, RAE can mitigate the effect of the device noise more efficiently as more samples are included in post-processing, improving the energy RMSE by two orders of magnitude.

\section{Conclusions}
\label{sec:concl}
This paper presented the first application of RAE to estimating the ground state energy of chemical Hamiltonians describing the hydrogen molecule in the minimum basis. 
RAE offers an efficient alternative to direct observable estimation, commonly used in VQE, 
significantly reducing the error of estimated expectation values despite 
relying on a simple depolarizing noise model that does not capture the complexity of decoherence effects in realistic quantum devices.
Additionally, RAE offers a more efficient way to benchmark device quality than straightforward VQE, albeit with the caveat that the underlying noise model may bias the benchmark toward certain device architectures. While the extracted noise parameter is not always a reliable indicator of device noise and can be imprecise, the convergence of the energy and its standard deviation with increasing RAE circuit layers provides an indicator of device quality without the need to compute reference data.

In contrast to previous RAE experiments that used randomized compiling to mitigate coherent error~\cite{dalal_noise_2023},
we took a simpler approach by analyzing the likelihood functions for a series of IBM devices. We selected the QPU whose noise model was overall consistent with the depolarizing noise assumption, allowing us to study the impact of other factors, such as device parameter instability, on RAE performance. In particular, {\it ibmq\_montreal} was found to exhibit both the lowest coherent error and the most stable (reproducible) effective noise parameter and was used for subsequent RAE experiments. 

For the one-qubit version of the ground state energy estimation problem, the effective noise parameter was in the interval of $2\cdot 10^{-3} - 3\cdot 10^{-3}$ on {\it ibmq\_montreal}. Despite low noise levels, RAE performance did not always improve as $L_{max}$ increased, indicating the sensitivity of the method to the coherent error and discrepancies between ideal and realistic noise description. Nevertheless, the fast convergence of the total energy RMSE obtained from experimental data offered preliminary evidence of RAE advantage as an estimation technique. 

In the two-qubit implementation of the ground state energy estimation, we were more limited in circuit depth due to increased noise levels, with the effective noise parameter estimated to be approximately $0.045 - 0.05$. Similar to the one-qubit case, we observed fast convergence of the total energy RMSE with RAE, resulting in the error reduction from 30 mHa (in direct sampling) to approximately 0.1 mHa (best RAE estimate).

In both experiments, RAE performed better for the expectation values of the Pauli terms, which exhibited large biases in direct sampling, resulting in effective error reduction. In view of this noise mitigation property, an important direction of future work is to compare the results of RAE to those of error-mitigated VQE under a fixed shot budget. Error mitigation techniques such as zero-noise extrapolation~\cite{temme_error_2017,kandala_error_2019,krebsbach_optimization_2022}, probabilistic error cancellation~\cite{temme_error_2017,kandala_error_2019}, and Clifford data regression~\cite{czarnik_error_2021,lowe_unified_2021} can reduce the bias of estimated expectation values in hardware experiments at the expense of allocating additional samples to infer the relevant aspects of device noise. Under a fixed measurement budget, however, this may increase sampling noise (i.e., the standard deviation of the estimator). The problem is further aggravated when several techniques are combined in an error mitigation pipeline~\cite{saki_hypothesis_2023}. In contrast, RAE can optimize quantum resource usage by tuning the layer schedule and shot allocation subject to constraints, yielding an overall improvement of RMSE~\cite{giurgica-tiron_low_2022}.

A major limitation of the near-term RAE is its incompatibility with grouping techniques, precluding faithful comparisons to the most advanced observable estimation routines applied in VQE. Fault-tolerant computation protocols can overcome this restriction. In particular, advances in Hamiltonian encoding~\cite{loaiza_reducing_2023, loaiza_block-invariant_2023, loaiza_majorana_2024, rocca_reducing_2024} could unlock new applications of RAE in the emerging era of early fault-tolerant quantum computing~\cite{katabarwa_early_2024}. Exploring corresponding algorithmic trade-offs is an important direction of future research, paving the way to further enhancing the utility of RAE in quantum chemistry.

\section*{Acknowledgements \label{Acknowledgement}} 
The authors acknowledge insightful discussions and suggestions from Maxwell Radin, Peter Johnson, Yanbing Zhou, and Amara Katabarwa. Numerical results were generated using the Orquestra\textregistered{} platform by Zapata Computing, Inc.

\newpage
\appendix
\setcounter{table}{0}
\setcounter{figure}{0}
\renewcommand{\thetable}{\thesection\arabic{table}}
\renewcommand{\thefigure}{\thesection\arabic{figure}} 

\section{Noisy Chebyshev likelihood functions}
\label{app:likelihood_fit}
\begin{figure}[!h]
\center
\includegraphics[width=0.9\linewidth]{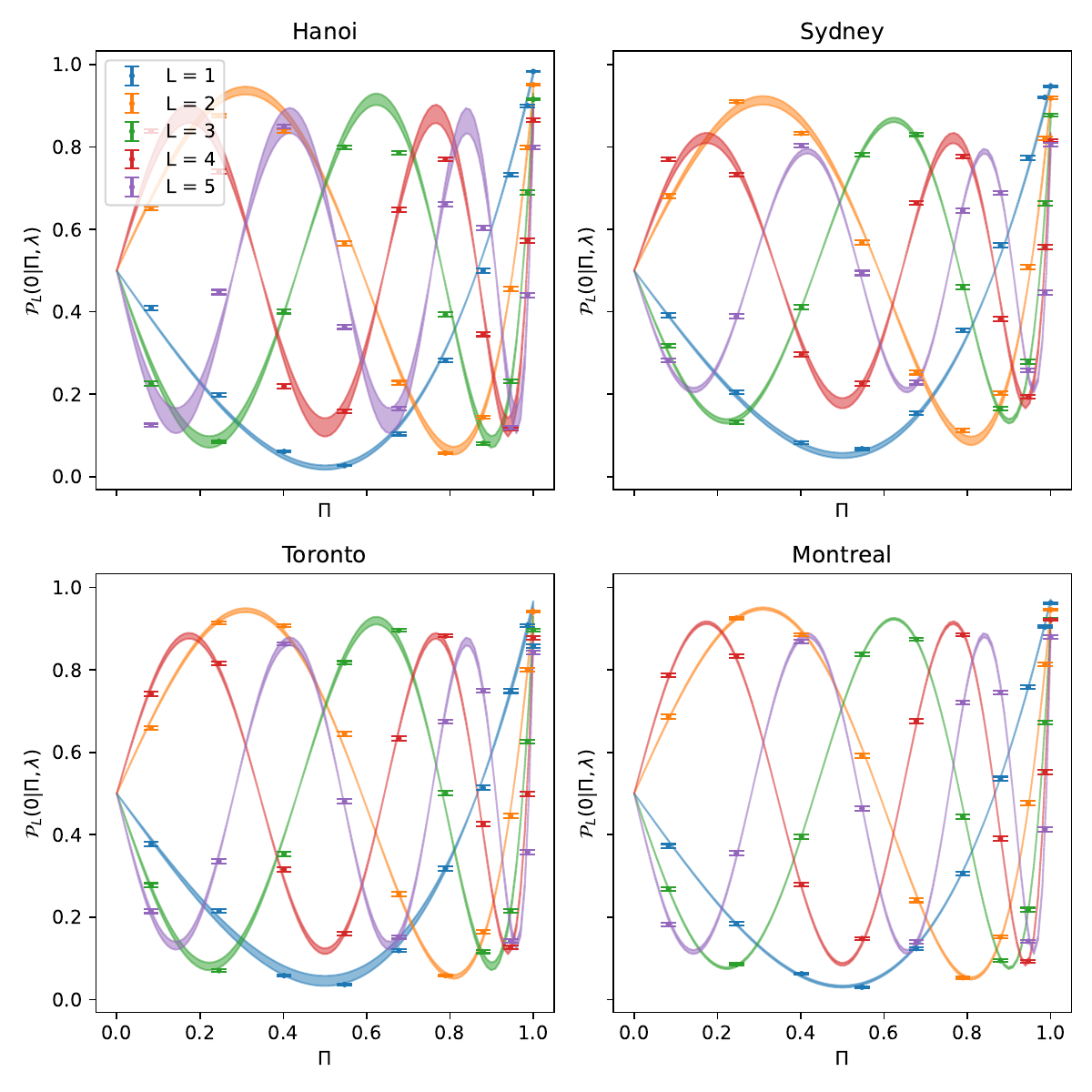}
\caption{The likelihood of obtaining even bitstrings when measuring $X_0X_1$ for RAE circuits with $L=1,2,3,4,5$ on premium IBM devices. Solid lines represent least squares fits to experimentally estimated likelihood values.} 
\label{fig:rae_likelihood_premium}
\end{figure}

Chebyshev likelihood functions fitted to experimental data for \textit{ibmq\_hanoi}, \textit{ibmq\_sydney}, \textit{ibmq\_toronto}, and \textit{ibmq\_montreal} are presented in Fig.~\ref{fig:rae_likelihood_premium}. Noise parameters corresponding to the solid lines can be found in Tab.~\ref{tab:lambda_eff_premium} along with the error bars. 

\begin{table}
\center
\begin{tabular}{lccccc}
\hline
%\hline
backend/L & 1 & 2 & 3 & 4 & 5 \\
\hline
\hline
Hanoi & 0.030(7) & 0.054(9) & 0.053(10) & 0.061(13) & 0.058(15) \\
Montreal & 0.043(4) & 0.043(3) & 0.047(2) & 0.042(2) & 0.048(2) \\
Sydney & 0.072(9) & 0.076(10) & 0.089(4) & 0.098(9) & 0.099(4) \\
Toronto & 0.063(18) & 0.047(4) & 0.050(6) & 0.059(4) & 0.055(5) \\
\hline
\end{tabular}
\caption{Effective noise parameters $\lambda$ from fitted likelihood functions shown in Fig.~\ref{fig:rae_likelihood_premium} }\label{tab:lambda_eff_premium}
\end{table}

\section{Analysis of the likelihood functions for the one-qubit RAE}
\label{app:ll_1q}

To extract expectation values from the RAE experiment data, we computed the log-likelihood functions $l(\mathbf{d}; \Pi)$ for $X_0$ and $Y_0$ terms in the one-qubit molecular Hamiltonian on the uniform grids having 100 and 10000 points along $\lambda$ and $\Pi$ axes, respectively.  As a part of the analysis, we identified the optimal inference intervals for $\lambda$ and $\Pi$ that were later used to obtain numerical estimates of the expectation values.

\begin{figure}
\center
\includegraphics[width=0.9\linewidth]{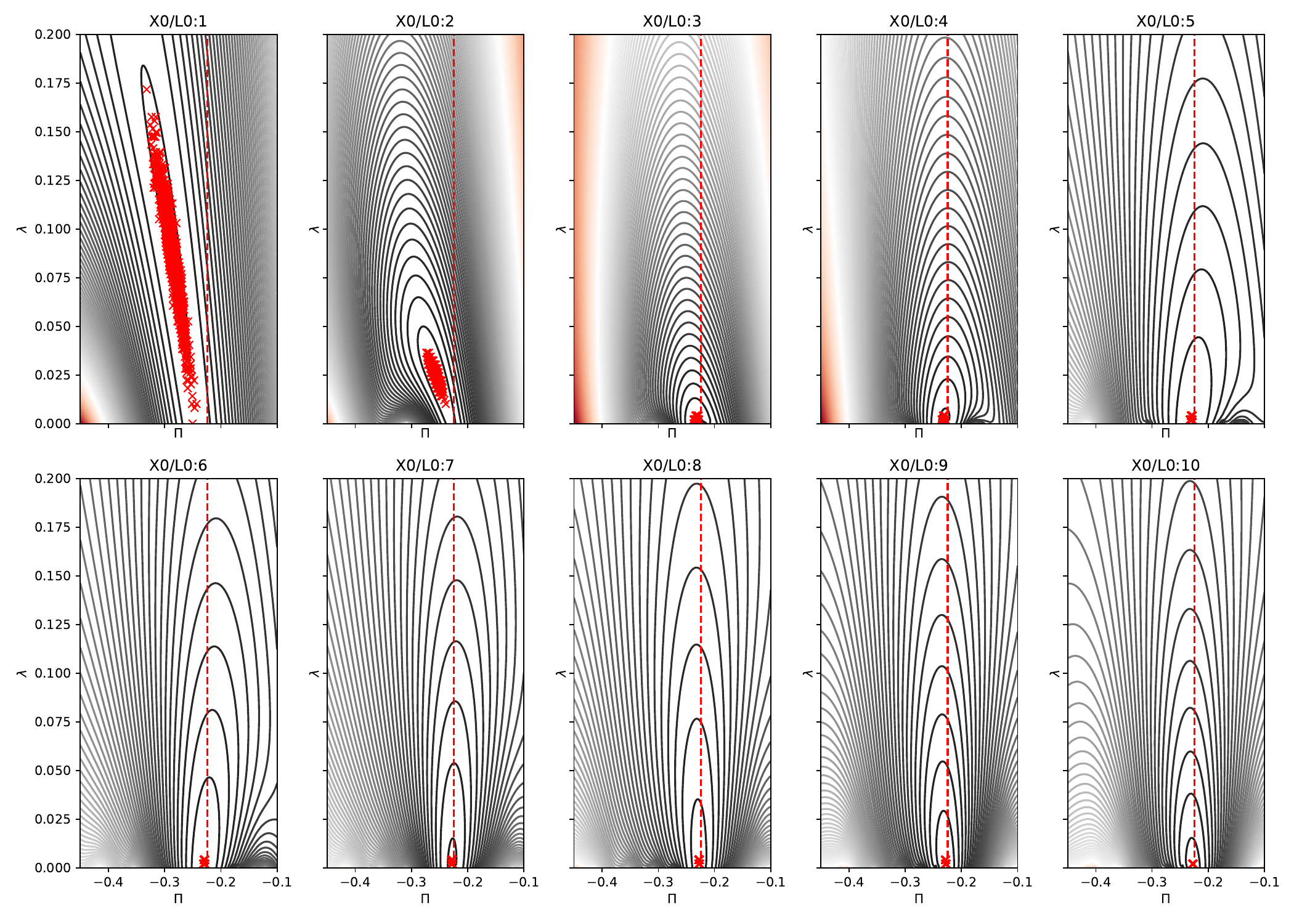}
\caption{Log-likelihood functions for $X_0$ generated from RAE experimental data on  \textit{ibmq\_montreal}. The vertical dashed line shows the reference expectation value. Red crosses mark the MLE from 1000 bootstrap runs. L0:X notation indicates that we included samples from RAE circuits with 0, 1, ..., X layers when calculating the likelihood function. Darker colors correspond to larger values of $l(\mathbf{d}; \Pi)$.} 
\label{fig:rae_logL_x01q_montreal}
\end{figure}

\begin{figure}
\center
\includegraphics[width=0.75\linewidth]{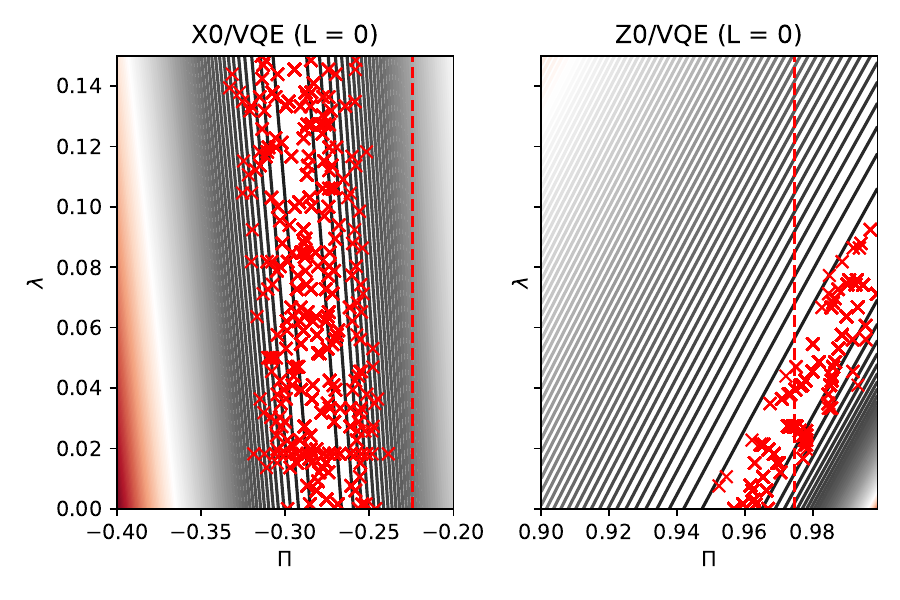}
\caption{Log-likelihood functions for $X_0$, $Z_0$ generated from the RAE  experiment data on \textit{ibmq\_montreal} for $L = 0$. Vertical dashed lines mark exact (noiseless) values. Red crosses represent the MLE from 5000 bootstrap runs. Darker colors correspond to larger values of the log-likelihood.} 
\label{fig:vqe_posterior_x0z01q_montreal}
\end{figure}

Unique maxima were identified for all the likelihood functions. In Fig.~\ref{fig:rae_logL_x01q_montreal}, we show a representative plot of $l(\mathbf{d}; \Pi)$ for $X_0$ on \textit{ibmq\_montreal}. As the maximum number of RAE layers increased from 1 to 10, the bootstrap MLE estimates (red crosses in Fig.~\ref{fig:rae_logL_x01q_montreal}) of $\langle X_0 \rangle$ approached the exact expectation value (marked with the vertical dashed line). 
The appearance of the log-likelihood functions for $L = 0$ and $L \ne 0$ is qualitatively different. As expected from Eq.~\ref{eq:cheb_likelihood}, maximum likelihood estimates of the Pauli expectations for the former are not well defined, i.e., the likelihood functions have degenerate maxima, making it impossible to identify a unique $(\Pi, \lambda)$ pair as shown in Fig.~\ref{fig:vqe_posterior_x0z01q_montreal}.

For this reason, $\lambda$ needs to be eliminated from MLE to obtain unique estimates of Pauli expectations for $L = 0$. To this end, we set it to 0 when extracting the expectation values. It can be shown that such approach is equivalent to the direct sampling used in standard VQE~\footnote{Indeed, the log-likelihood function in this case is $l(\mathbf{d}; \Pi) \propto \log(1-\Pi)(N_s - e_0) + \log(1+\Pi) e_0$, and the MLE estimate of $\Pi$ is $\hat{\Pi} = \frac{2e_0 - N_s}{N_s}$}. The ground state energies estimated from the individual Pauli expectations with this method are reported in Table~\ref{tab:vqe_1q_ibm_premium}.

\begin{table}
\center
\begin{tabular}{lcccc}
\hline
\hline
backend &  Hanoi & Sydney &  Toronto & Montreal \\
\hline
E, Ha & -1.1373(27)  & -1.1370(24)  &  -1.0451(51)  &  -1.1348(32)  \\
RMSE, mHa &  3  & 3 & 92 & 4  \\
\hline
\end{tabular}
\caption{Ground state energy estimates and their RMSE for the one-qubit problem calculated via MLE with $\lambda = 0$ (i.e., standard sampling) and 25000 bootstrap samples}\label{tab:vqe_1q_ibm_premium}
\end{table}

The ground state energies are close to the noiseless reference, consistent with generally low noise levels in one-qubit experiments, except for \textit{ibmq\_toronto}.

\section{Analysis of the likelihood functions for the two-qubit RAE}
\label{app:ll_2q}

\begin{figure}
\center
\includegraphics[scale=0.5]{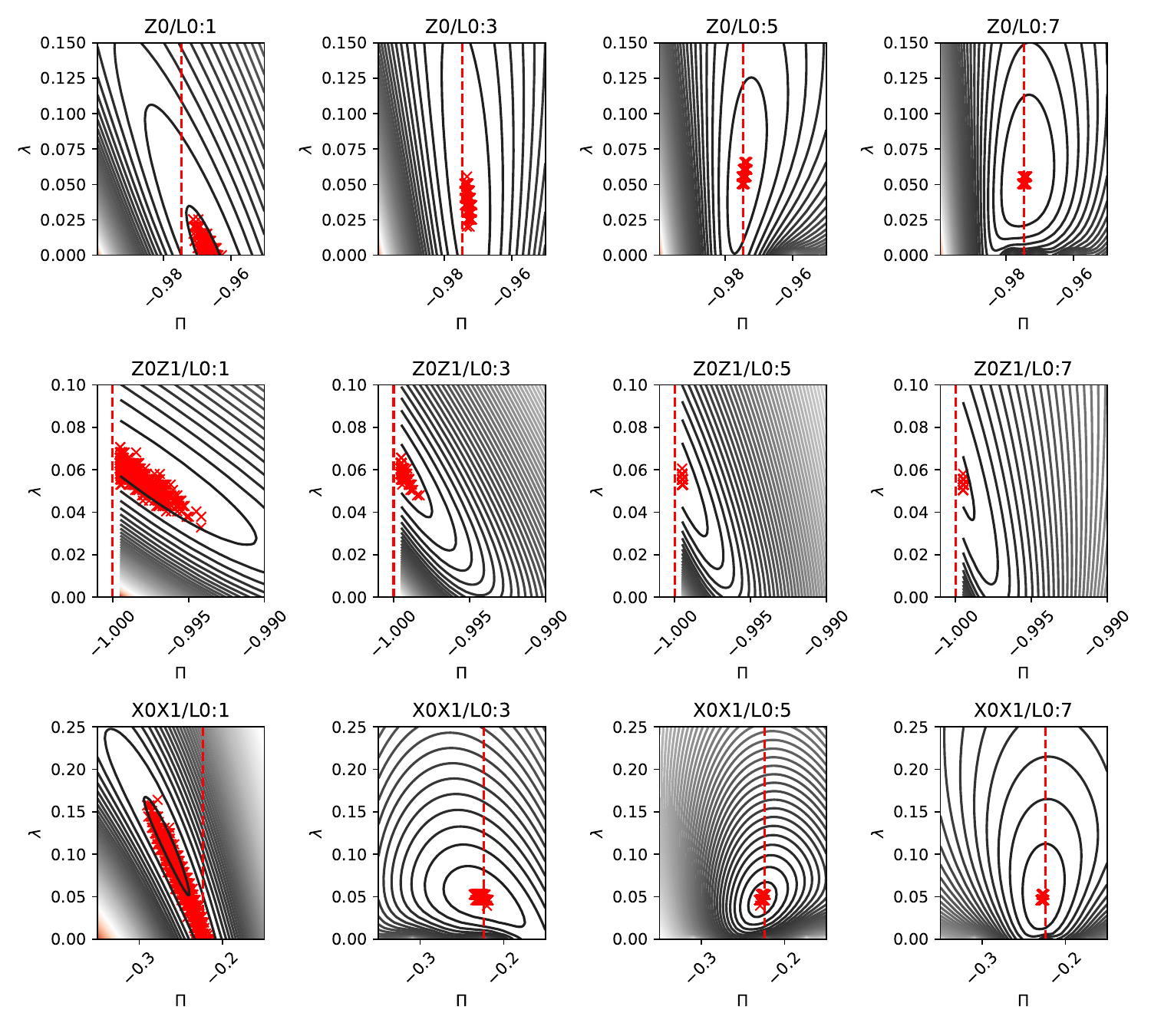}
\caption{Log-likelihood functions for $Z_0, Z_0Z_1$ and $X_0X_1$ on \textit{ibmq\_montreal}. Vertical dashed lines mark exact (noiseless) expectation values. Red crosses represent the MLE from 1000 bootstrap samples. Darker lines correspond to larger values of the log-likelihood.} 
\label{fig:rae_logL_multi2q_montreal}
\end{figure}

Compared to the one-qubit problem, the two-qubit expectation value estimation is more challenging due to the faster accumulation of coherent error with increased Grover depth. When analyzing the results of the two-qubit RAE experiments, we identified different convergence patterns exemplified in Fig.~\ref{fig:rae_logL_multi2q_montreal} for \textit{ibmq\_montreal}. Although all the estimates tend to stabilize as the maximum number of layers is increased from 1 to 8, some may have sizable fluctuations deviating from the exact value as $L_{max}$ approaches 8. Likewise, the maximum of the log-likelihood shifts along the $\lambda$ axis as $L_{max}$ increases for $Z_0$, which can be related to device parameter instability. 

As explained in Appendix~\ref{app:ll_1q}, MLE cannot be directly applied to the set of samples for $L = 0$ unless the nuisance parameter $\lambda$ is eliminated from the inference. Using this approach, we obtained the estimates of the ground state energy, reported in Tab.~\ref{tab:vqe_2q_ibm_premium}

\begin{table}
\center
\begin{tabular}{lcccc}
\hline
\hline
backend &  Hanoi & Sydney &  Toronto & Montreal \\
\hline
E, Ha &       -1.1047(29) & -1.0889(31) & -1.1156(26) &  -1.11282(29) \\
RMSE, mHa & 41  & 57 & 30  & 33  \\
\hline
\end{tabular}
\caption{Reference VQE energies and their RMSE for the 2-qubit ground state energy estimation problem}\label{tab:vqe_2q_ibm_premium}
\end{table}

\section{Noise Robust Incremental Sequence}
\label{app:schedules}

\begin{figure}[!h]
\center
\includegraphics[width=0.4\linewidth]{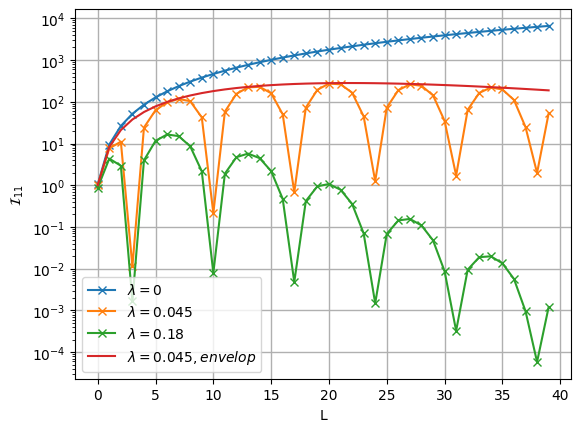}
\caption{Fisher information of $X_0 X_1$ where $\langle X_0 X_1 \rangle \approx -0.2237$ versus number of Grover layers $L$, under effective noise parameter $\lambda = 0, 0.045, 0.18$.} 
\label{fig:noise-incremental-sequence}
\end{figure}

In this Appendix, we discuss the method we used to optimize incremental sequence in the presence of noise. It can be viewed as an adaptation of the approach presented by Giurgica-Tiron et al.~\cite{giurgica-tiron_low_2022} to the problem of optimal shot allocation based on maximizing noisy Fisher information. Specifically, we explored a setting where the number of shots per RAE circuit is kept fixed while the Grover depth is subject to optimization. As one can see in Fig.~\ref{fig:noise-incremental-sequence}, while the Fisher information grows with the number of Grover layers in the absence of decoherence effects, i.e., $\lambda = 0$, it behaves differently when effective noise $\lambda >0$. The latter features an oscillation pattern with a period of approximately 7. It is desirable to design a layer schedule such that the local minima of Fisher information are avoided. This is achieved for the values of $L$ satisfying an approximate condition: 

\begin{align}
    \label{eq:noise-period}
    \operatorname{sin}^2{\left(\left(2L +1\right) \operatorname{acos}{\left(\Pi \right)}\right)} \approx 1, L\in \mathbb{Z}.
\end{align}
For the edge case where $\Pi \approx \pm 1$, the above condition is hard to satisfy, and the exponential incremental sequence is optimal, which coincides with the noiseless situation.

The oscillation has an envelope with a maximum at a certain number of Grover layers for noisy situations, whose analytical form can be derived by substituting condition \ref{eq:noise-period} to the Fisher information (Eq.~\ref{eq:fisher_info11}): 
\begin{align}
    \label{eq:noise-stop}
    \mathcal{I}(\Pi, \lambda)_{11, \text{envelop}} = \frac{\left(2 L + 1\right)^{2} e^{- \lambda \left(2L + 1\right)}}{(1-\Pi^{2})}.
\end{align}
The number of Grover layers $L_{\text{max}\mathcal{I}}$ that maximizes $\mathcal{I}(\Pi, \lambda)_{11, \text{envelop}}$ is the only root of equation $\frac{\partial \mathcal{I}(\Pi, \lambda)_{11, \text{envelop}}}{\partial L} = 0$, which yields $L_{\text{max}\mathcal{I}} = \frac{1}{\lambda} +\frac{1}{2}$. This result is consistent with Fig.~\ref{fig:noise-incremental-sequence} where Fisher information reaches maxima at  $L_{\text{max}\mathcal{I}}\approx 22$ and 5 for $\lambda = 0.045$ and 0.18, respectively.

Based on the discussion above, we designed our noise robust incremental sequence as follows: 
\begin{itemize}
    \item Measure $\lambda$ and calculate $L_{\text{max}\mathcal{I}}$; choose a hyperparameter $c>0$
    \item If $|\Pi|<c\lambda$ or $1-|\Pi|<c\lambda$, do exponential incremental sequence.
    \item Otherwise, the incremental sequence is constructed by $L$ that satisfy:
    \begin{itemize}
        \item[$\blacksquare$] $\operatorname{sin}^2{\left(\left(2L +1\right) \operatorname{acos}{\left(\Pi \right)}\right)} > 1-c\lambda $
        \item[$\blacksquare$] $L\in \mathbb{Z}$
        \item[$\blacksquare$] $L < L_{\text{max}\mathcal{I}}$
\end{itemize}
\end{itemize}

\begin{figure}[!h]
\center
\includegraphics[width=0.6\linewidth]{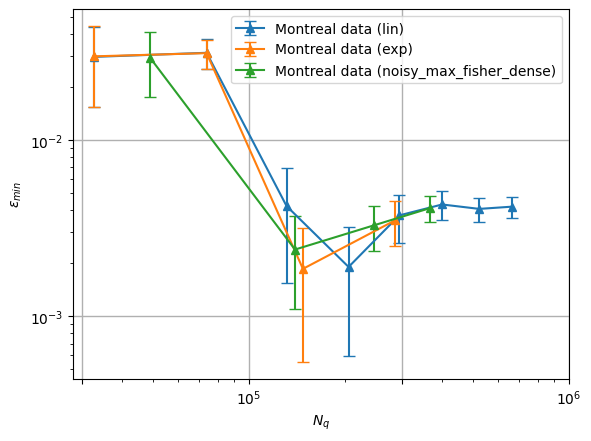}
\caption{The comparison between incremental sequences (Linear, Exponential, and Noise robust) for Montreal data.} 
\label{fig:incremental-sequences-compare}
\end{figure}

We compared the performance of different incremental sequences in Fig.~\ref{fig:incremental-sequences-compare}. Our noise-robust incremental sequence reaches minimum error with fewer queries than linear and exponential ones. However, the minimum error reached is larger compared to the other two sequences. This could be due to our model assumptions regarding device noise that ignore the complexity of decoherence effects on real hardware.

\FloatBarrier


\begin{thebibliography}{49}%
\makeatletter
\providecommand \@ifxundefined [1]{%
 \@ifx{#1\undefined}
}%
\providecommand \@ifnum [1]{%
 \ifnum #1\expandafter \@firstoftwo
 \else \expandafter \@secondoftwo
 \fi
}%
\providecommand \@ifx [1]{%
 \ifx #1\expandafter \@firstoftwo
 \else \expandafter \@secondoftwo
 \fi
}%
\providecommand \natexlab [1]{#1}%
\providecommand \enquote  [1]{``#1''}%
\providecommand \bibnamefont  [1]{#1}%
\providecommand \bibfnamefont [1]{#1}%
\providecommand \citenamefont [1]{#1}%
\providecommand \href@noop [0]{\@secondoftwo}%
\providecommand \href [0]{\begingroup \@sanitize@url \@href}%
\providecommand \@href[1]{\@@startlink{#1}\@@href}%
\providecommand \@@href[1]{\endgroup#1\@@endlink}%
\providecommand \@sanitize@url [0]{\catcode `\\12\catcode `\$12\catcode
  `\&12\catcode `\#12\catcode `\^12\catcode `\_12\catcode `\%12\relax}%
\providecommand \@@startlink[1]{}%
\providecommand \@@endlink[0]{}%
\providecommand \url  [0]{\begingroup\@sanitize@url \@url }%
\providecommand \@url [1]{\endgroup\@href {#1}{\urlprefix }}%
\providecommand \urlprefix  [0]{URL }%
\providecommand \Eprint [0]{\href }%
\providecommand \doibase [0]{https://doi.org/}%
\providecommand \selectlanguage [0]{\@gobble}%
\providecommand \bibinfo  [0]{\@secondoftwo}%
\providecommand \bibfield  [0]{\@secondoftwo}%
\providecommand \translation [1]{[#1]}%
\providecommand \BibitemOpen [0]{}%
\providecommand \bibitemStop [0]{}%
\providecommand \bibitemNoStop [0]{.\EOS\space}%
\providecommand \EOS [0]{\spacefactor3000\relax}%
\providecommand \BibitemShut  [1]{\csname bibitem#1\endcsname}%
\let\auto@bib@innerbib\@empty
%</preamble>
\bibitem [{\citenamefont {Dalzell}\ \emph {et~al.}(2023)\citenamefont
  {Dalzell}, \citenamefont {McArdle}, \citenamefont {Berta}, \citenamefont
  {Bienias}, \citenamefont {Chen}, \citenamefont {Gilyén}, \citenamefont
  {Hann}, \citenamefont {Kastoryano}, \citenamefont {Khabiboulline},
  \citenamefont {Kubica}, \citenamefont {Salton}, \citenamefont {Wang},\ and\
  \citenamefont {Brandão}}]{dalzell_quantum_2023}%
  \BibitemOpen
  \bibfield  {author} {\bibinfo {author} {\bibfnamefont {A.~M.}\ \bibnamefont
  {Dalzell}}, \bibinfo {author} {\bibfnamefont {S.}~\bibnamefont {McArdle}},
  \bibinfo {author} {\bibfnamefont {M.}~\bibnamefont {Berta}}, \bibinfo
  {author} {\bibfnamefont {P.}~\bibnamefont {Bienias}}, \bibinfo {author}
  {\bibfnamefont {C.-F.}\ \bibnamefont {Chen}}, \bibinfo {author}
  {\bibfnamefont {A.}~\bibnamefont {Gilyén}}, \bibinfo {author} {\bibfnamefont
  {C.~T.}\ \bibnamefont {Hann}}, \bibinfo {author} {\bibfnamefont {M.~J.}\
  \bibnamefont {Kastoryano}}, \bibinfo {author} {\bibfnamefont {E.~T.}\
  \bibnamefont {Khabiboulline}}, \bibinfo {author} {\bibfnamefont
  {A.}~\bibnamefont {Kubica}}, \bibinfo {author} {\bibfnamefont
  {G.}~\bibnamefont {Salton}}, \bibinfo {author} {\bibfnamefont
  {S.}~\bibnamefont {Wang}},\ and\ \bibinfo {author} {\bibfnamefont {F.~G.
  S.~L.}\ \bibnamefont {Brandão}},\ }\href
  {https://doi.org/10.48550/arXiv.2310.03011} {\bibinfo {title} {Quantum
  algorithms: {A} survey of applications and end-to-end complexities}}
  (\bibinfo {year} {2023}),\ \bibinfo {note} {arXiv:2310.03011
  [quant-ph]}\BibitemShut {NoStop}%
\bibitem [{\citenamefont {Wang}\ \emph {et~al.}(2021)\citenamefont {Wang},
  \citenamefont {Koh}, \citenamefont {Johnson},\ and\ \citenamefont
  {Cao}}]{wang2020bayesian}%
  \BibitemOpen
  \bibfield  {author} {\bibinfo {author} {\bibfnamefont {G.}~\bibnamefont
  {Wang}}, \bibinfo {author} {\bibfnamefont {D.~E.}\ \bibnamefont {Koh}},
  \bibinfo {author} {\bibfnamefont {P.~D.}\ \bibnamefont {Johnson}},\ and\
  \bibinfo {author} {\bibfnamefont {Y.}~\bibnamefont {Cao}},\ }\bibfield
  {title} {\bibinfo {title} {Minimizing estimation runtime on noisy quantum
  computers},\ }\href {https://doi.org/10.1103/PRXQuantum.2.010346} {\bibfield
  {journal} {\bibinfo  {journal} {PRX Quantum}\ }\textbf {\bibinfo {volume}
  {2}},\ \bibinfo {pages} {010346} (\bibinfo {year} {2021})}\BibitemShut
  {NoStop}%
\bibitem [{\citenamefont {Koh}\ \emph {et~al.}()\citenamefont {Koh},
  \citenamefont {Wang}, \citenamefont {Johnson},\ and\ \citenamefont
  {Cao}}]{ELF}%
  \BibitemOpen
  \bibfield  {author} {\bibinfo {author} {\bibfnamefont {D.~E.}\ \bibnamefont
  {Koh}}, \bibinfo {author} {\bibfnamefont {G.}~\bibnamefont {Wang}}, \bibinfo
  {author} {\bibfnamefont {P.~D.}\ \bibnamefont {Johnson}},\ and\ \bibinfo
  {author} {\bibfnamefont {Y.}~\bibnamefont {Cao}},\ }\href@noop {} {\bibinfo
  {title} {A framework for engineering quantum likelihood functions for
  expectation estimation}},\ \Eprint {https://arxiv.org/abs/arXiv:2006.09349v1}
  {arXiv:2006.09349v1} \BibitemShut {NoStop}%
\bibitem [{\citenamefont {Suzuki}\ \emph {et~al.}(2020)\citenamefont {Suzuki},
  \citenamefont {Uno}, \citenamefont {Raymond}, \citenamefont {Tanaka},
  \citenamefont {Onodera},\ and\ \citenamefont
  {Yamamoto}}]{suzuki_amplitude_2020}%
  \BibitemOpen
  \bibfield  {author} {\bibinfo {author} {\bibfnamefont {Y.}~\bibnamefont
  {Suzuki}}, \bibinfo {author} {\bibfnamefont {S.}~\bibnamefont {Uno}},
  \bibinfo {author} {\bibfnamefont {R.}~\bibnamefont {Raymond}}, \bibinfo
  {author} {\bibfnamefont {T.}~\bibnamefont {Tanaka}}, \bibinfo {author}
  {\bibfnamefont {T.}~\bibnamefont {Onodera}},\ and\ \bibinfo {author}
  {\bibfnamefont {N.}~\bibnamefont {Yamamoto}},\ }\bibfield  {title} {\bibinfo
  {title} {Amplitude estimation without phase estimation},\ }\href
  {https://doi.org/10.1007/s11128-019-2565-2} {\bibfield  {journal} {\bibinfo
  {journal} {Quantum Information Processing}\ }\textbf {\bibinfo {volume}
  {19}},\ \bibinfo {pages} {75} (\bibinfo {year} {2020})},\ \bibinfo {note}
  {arXiv:1904.10246 [quant-ph]}\BibitemShut {NoStop}%
\bibitem [{\citenamefont {Tanaka}\ \emph {et~al.}(2021)\citenamefont {Tanaka},
  \citenamefont {Suzuki}, \citenamefont {Uno}, \citenamefont {Raymond},
  \citenamefont {Onodera},\ and\ \citenamefont
  {Yamamoto}}]{tanaka_amplitude_2021}%
  \BibitemOpen
  \bibfield  {author} {\bibinfo {author} {\bibfnamefont {T.}~\bibnamefont
  {Tanaka}}, \bibinfo {author} {\bibfnamefont {Y.}~\bibnamefont {Suzuki}},
  \bibinfo {author} {\bibfnamefont {S.}~\bibnamefont {Uno}}, \bibinfo {author}
  {\bibfnamefont {R.}~\bibnamefont {Raymond}}, \bibinfo {author} {\bibfnamefont
  {T.}~\bibnamefont {Onodera}},\ and\ \bibinfo {author} {\bibfnamefont
  {N.}~\bibnamefont {Yamamoto}},\ }\bibfield  {title} {\bibinfo {title}
  {Amplitude estimation via maximum likelihood on noisy quantum computer},\
  }\href {https://doi.org/10.1007/s11128-021-03215-9} {\bibfield  {journal}
  {\bibinfo  {journal} {Quantum Information Processing}\ }\textbf {\bibinfo
  {volume} {20}},\ \bibinfo {pages} {293} (\bibinfo {year} {2021})}\BibitemShut
  {NoStop}%
\bibitem [{\citenamefont {Gonthier}\ \emph {et~al.}(2022)\citenamefont
  {Gonthier}, \citenamefont {Radin}, \citenamefont {Buda}, \citenamefont
  {Doskocil}, \citenamefont {Abuan},\ and\ \citenamefont
  {Romero}}]{gonthier_measurements_2022}%
  \BibitemOpen
  \bibfield  {author} {\bibinfo {author} {\bibfnamefont {J.~F.}\ \bibnamefont
  {Gonthier}}, \bibinfo {author} {\bibfnamefont {M.~D.}\ \bibnamefont {Radin}},
  \bibinfo {author} {\bibfnamefont {C.}~\bibnamefont {Buda}}, \bibinfo {author}
  {\bibfnamefont {E.~J.}\ \bibnamefont {Doskocil}}, \bibinfo {author}
  {\bibfnamefont {C.~M.}\ \bibnamefont {Abuan}},\ and\ \bibinfo {author}
  {\bibfnamefont {J.}~\bibnamefont {Romero}},\ }\bibfield  {title} {\bibinfo
  {title} {Measurements as a roadblock to near-term practical quantum advantage
  in chemistry: {Resource} analysis},\ }\href
  {https://doi.org/10.1103/PhysRevResearch.4.033154} {\bibfield  {journal}
  {\bibinfo  {journal} {Physical Review Research}\ }\textbf {\bibinfo {volume}
  {4}},\ \bibinfo {pages} {033154} (\bibinfo {year} {2022})}\BibitemShut
  {NoStop}%
\bibitem [{\citenamefont {Peruzzo}\ \emph {et~al.}(2014)\citenamefont
  {Peruzzo}, \citenamefont {McClean}, \citenamefont {Shadbolt}, \citenamefont
  {Yung}, \citenamefont {Zhou}, \citenamefont {Love}, \citenamefont
  {Aspuru-Guzik},\ and\ \citenamefont {O$'$brien}}]{Peruzzo2014}%
  \BibitemOpen
  \bibfield  {author} {\bibinfo {author} {\bibfnamefont {A.}~\bibnamefont
  {Peruzzo}}, \bibinfo {author} {\bibfnamefont {J.}~\bibnamefont {McClean}},
  \bibinfo {author} {\bibfnamefont {P.}~\bibnamefont {Shadbolt}}, \bibinfo
  {author} {\bibfnamefont {M.-H.}\ \bibnamefont {Yung}}, \bibinfo {author}
  {\bibfnamefont {X.-Q.}\ \bibnamefont {Zhou}}, \bibinfo {author}
  {\bibfnamefont {P.~J.}\ \bibnamefont {Love}}, \bibinfo {author}
  {\bibfnamefont {A.}~\bibnamefont {Aspuru-Guzik}},\ and\ \bibinfo {author}
  {\bibfnamefont {J.~L.}\ \bibnamefont {O$'$brien}},\ }\bibfield  {title}
  {\bibinfo {title} {A variational eigenvalue solver on a photonic quantum
  processor},\ }\href@noop {} {\bibfield  {journal} {\bibinfo  {journal} {Nat.
  Commun.}\ }\textbf {\bibinfo {volume} {5}},\ \bibinfo {pages} {4213}
  (\bibinfo {year} {2014})}\BibitemShut {NoStop}%
\bibitem [{\citenamefont {Johnson}\ \emph {et~al.}(2022)\citenamefont
  {Johnson}, \citenamefont {Kunitsa}, \citenamefont {Gonthier}, \citenamefont
  {Radin}, \citenamefont {Buda}, \citenamefont {Doskocil}, \citenamefont
  {Abuan},\ and\ \citenamefont {Romero}}]{johnson_reducing_2022}%
  \BibitemOpen
  \bibfield  {author} {\bibinfo {author} {\bibfnamefont {P.~D.}\ \bibnamefont
  {Johnson}}, \bibinfo {author} {\bibfnamefont {A.~A.}\ \bibnamefont
  {Kunitsa}}, \bibinfo {author} {\bibfnamefont {J.~F.}\ \bibnamefont
  {Gonthier}}, \bibinfo {author} {\bibfnamefont {M.~D.}\ \bibnamefont {Radin}},
  \bibinfo {author} {\bibfnamefont {C.}~\bibnamefont {Buda}}, \bibinfo {author}
  {\bibfnamefont {E.~J.}\ \bibnamefont {Doskocil}}, \bibinfo {author}
  {\bibfnamefont {C.~M.}\ \bibnamefont {Abuan}},\ and\ \bibinfo {author}
  {\bibfnamefont {J.}~\bibnamefont {Romero}},\ }\href
  {https://doi.org/10.48550/arXiv.2203.07275} {\bibinfo {title} {Reducing the
  cost of energy estimation in the variational quantum eigensolver algorithm
  with robust amplitude estimation}} (\bibinfo {year} {2022})\BibitemShut
  {NoStop}%
\bibitem [{\citenamefont {Oftelie}\ \emph {et~al.}(2021)\citenamefont
  {Oftelie}, \citenamefont {Urbanek}, \citenamefont {Metcalf}, \citenamefont
  {Carter}, \citenamefont {Kemper},\ and\ \citenamefont
  {Jong}}]{oftelie_simulating_2021}%
  \BibitemOpen
  \bibfield  {author} {\bibinfo {author} {\bibfnamefont {L.~B.}\ \bibnamefont
  {Oftelie}}, \bibinfo {author} {\bibfnamefont {M.}~\bibnamefont {Urbanek}},
  \bibinfo {author} {\bibfnamefont {M.}~\bibnamefont {Metcalf}}, \bibinfo
  {author} {\bibfnamefont {J.}~\bibnamefont {Carter}}, \bibinfo {author}
  {\bibfnamefont {A.~F.}\ \bibnamefont {Kemper}},\ and\ \bibinfo {author}
  {\bibfnamefont {W.~A.~d.}\ \bibnamefont {Jong}},\ }\bibfield  {title}
  {\bibinfo {title} {Simulating quantum materials with digital quantum
  computers},\ }\href {https://doi.org/10.1088/2058-9565/ac1ca6} {\bibfield
  {journal} {\bibinfo  {journal} {Quantum Science and Technology}\ }\textbf
  {\bibinfo {volume} {6}},\ \bibinfo {pages} {043002} (\bibinfo {year}
  {2021})}\BibitemShut {NoStop}%
\bibitem [{\citenamefont {Jaksch}\ \emph {et~al.}(2022)\citenamefont {Jaksch},
  \citenamefont {Givi}, \citenamefont {Daley},\ and\ \citenamefont
  {Rung}}]{jaksch_variational_2022}%
  \BibitemOpen
  \bibfield  {author} {\bibinfo {author} {\bibfnamefont {D.}~\bibnamefont
  {Jaksch}}, \bibinfo {author} {\bibfnamefont {P.}~\bibnamefont {Givi}},
  \bibinfo {author} {\bibfnamefont {A.~J.}\ \bibnamefont {Daley}},\ and\
  \bibinfo {author} {\bibfnamefont {T.}~\bibnamefont {Rung}},\ }\href
  {https://doi.org/10.48550/arXiv.2209.04915} {\bibinfo {title} {Variational
  {Quantum} {Algorithms} for {Computational} {Fluid} {Dynamics}}} (\bibinfo
  {year} {2022}),\ \bibinfo {note} {arXiv:2209.04915 [physics,
  physics:quant-ph]}\BibitemShut {NoStop}%
\bibitem [{\citenamefont {Leong}\ \emph {et~al.}(2022)\citenamefont {Leong},
  \citenamefont {Ewe},\ and\ \citenamefont {Koh}}]{leong_variational_2022}%
  \BibitemOpen
  \bibfield  {author} {\bibinfo {author} {\bibfnamefont {F.~Y.}\ \bibnamefont
  {Leong}}, \bibinfo {author} {\bibfnamefont {W.-B.}\ \bibnamefont {Ewe}},\
  and\ \bibinfo {author} {\bibfnamefont {D.~E.}\ \bibnamefont {Koh}},\
  }\bibfield  {title} {\bibinfo {title} {Variational {Quantum} {Evolution}
  {Equation} {Solver}},\ }\href {https://doi.org/10.1038/s41598-022-14906-3}
  {\bibfield  {journal} {\bibinfo  {journal} {Scientific Reports}\ }\textbf
  {\bibinfo {volume} {12}},\ \bibinfo {pages} {10817} (\bibinfo {year}
  {2022})},\ \bibinfo {note} {arXiv:2204.02912 [physics,
  physics:quant-ph]}\BibitemShut {NoStop}%
\bibitem [{\citenamefont {Hayakawa}(2022)}]{hayakawa_quantum_2022}%
  \BibitemOpen
  \bibfield  {author} {\bibinfo {author} {\bibfnamefont {R.}~\bibnamefont
  {Hayakawa}},\ }\bibfield  {title} {\bibinfo {title} {Quantum algorithm for
  persistent {Betti} numbers and topological data analysis},\ }\href
  {https://doi.org/10.22331/q-2022-12-07-873} {\bibfield  {journal} {\bibinfo
  {journal} {Quantum}\ }\textbf {\bibinfo {volume} {6}},\ \bibinfo {pages}
  {873} (\bibinfo {year} {2022})}\BibitemShut {NoStop}%
\bibitem [{\citenamefont {Berry}\ \emph {et~al.}(2024)\citenamefont {Berry},
  \citenamefont {Su}, \citenamefont {Gyurik}, \citenamefont {King},
  \citenamefont {Basso}, \citenamefont {Barba}, \citenamefont {Rajput},
  \citenamefont {Wiebe}, \citenamefont {Dunjko},\ and\ \citenamefont
  {Babbush}}]{berry_analyzing_2024}%
  \BibitemOpen
  \bibfield  {author} {\bibinfo {author} {\bibfnamefont {D.~W.}\ \bibnamefont
  {Berry}}, \bibinfo {author} {\bibfnamefont {Y.}~\bibnamefont {Su}}, \bibinfo
  {author} {\bibfnamefont {C.}~\bibnamefont {Gyurik}}, \bibinfo {author}
  {\bibfnamefont {R.}~\bibnamefont {King}}, \bibinfo {author} {\bibfnamefont
  {J.}~\bibnamefont {Basso}}, \bibinfo {author} {\bibfnamefont {A.~D.~T.}\
  \bibnamefont {Barba}}, \bibinfo {author} {\bibfnamefont {A.}~\bibnamefont
  {Rajput}}, \bibinfo {author} {\bibfnamefont {N.}~\bibnamefont {Wiebe}},
  \bibinfo {author} {\bibfnamefont {V.}~\bibnamefont {Dunjko}},\ and\ \bibinfo
  {author} {\bibfnamefont {R.}~\bibnamefont {Babbush}},\ }\bibfield  {title}
  {\bibinfo {title} {Analyzing {Prospects} for {Quantum} {Advantage} in
  {Topological} {Data} {Analysis}},\ }\href
  {https://doi.org/10.1103/PRXQuantum.5.010319} {\bibfield  {journal} {\bibinfo
   {journal} {PRX Quantum}\ }\textbf {\bibinfo {volume} {5}},\ \bibinfo {pages}
  {010319} (\bibinfo {year} {2024})}\BibitemShut {NoStop}%
\bibitem [{\citenamefont {Giurgica-Tiron}\ \emph
  {et~al.}(2022{\natexlab{a}})\citenamefont {Giurgica-Tiron}, \citenamefont
  {Kerenidis}, \citenamefont {Labib}, \citenamefont {Prakash},\ and\
  \citenamefont {Zeng}}]{giurgica-tiron_low_2022}%
  \BibitemOpen
  \bibfield  {author} {\bibinfo {author} {\bibfnamefont {T.}~\bibnamefont
  {Giurgica-Tiron}}, \bibinfo {author} {\bibfnamefont {I.}~\bibnamefont
  {Kerenidis}}, \bibinfo {author} {\bibfnamefont {F.}~\bibnamefont {Labib}},
  \bibinfo {author} {\bibfnamefont {A.}~\bibnamefont {Prakash}},\ and\ \bibinfo
  {author} {\bibfnamefont {W.}~\bibnamefont {Zeng}},\ }\bibfield  {title}
  {\bibinfo {title} {Low depth algorithms for quantum amplitude estimation},\
  }\href {https://doi.org/10.22331/q-2022-06-27-745} {\bibfield  {journal}
  {\bibinfo  {journal} {Quantum}\ }\textbf {\bibinfo {volume} {6}},\ \bibinfo
  {pages} {745} (\bibinfo {year} {2022}{\natexlab{a}})},\ \bibinfo {note}
  {arXiv:2012.03348 [quant-ph]}\BibitemShut {NoStop}%
\bibitem [{\citenamefont {Giurgica-Tiron}\ \emph
  {et~al.}(2022{\natexlab{b}})\citenamefont {Giurgica-Tiron}, \citenamefont
  {Johri}, \citenamefont {Kerenidis}, \citenamefont {Nguyen}, \citenamefont
  {Pisenti}, \citenamefont {Prakash}, \citenamefont {Sosnova}, \citenamefont
  {Wright},\ and\ \citenamefont {Zeng}}]{giurgica-tiron_low-depth_2022}%
  \BibitemOpen
  \bibfield  {author} {\bibinfo {author} {\bibfnamefont {T.}~\bibnamefont
  {Giurgica-Tiron}}, \bibinfo {author} {\bibfnamefont {S.}~\bibnamefont
  {Johri}}, \bibinfo {author} {\bibfnamefont {I.}~\bibnamefont {Kerenidis}},
  \bibinfo {author} {\bibfnamefont {J.}~\bibnamefont {Nguyen}}, \bibinfo
  {author} {\bibfnamefont {N.}~\bibnamefont {Pisenti}}, \bibinfo {author}
  {\bibfnamefont {A.}~\bibnamefont {Prakash}}, \bibinfo {author} {\bibfnamefont
  {K.}~\bibnamefont {Sosnova}}, \bibinfo {author} {\bibfnamefont
  {K.}~\bibnamefont {Wright}},\ and\ \bibinfo {author} {\bibfnamefont
  {W.}~\bibnamefont {Zeng}},\ }\bibfield  {title} {\bibinfo {title} {Low-depth
  amplitude estimation on a trapped-ion quantum computer},\ }\href
  {https://doi.org/10.1103/PhysRevResearch.4.033034} {\bibfield  {journal}
  {\bibinfo  {journal} {Physical Review Research}\ }\textbf {\bibinfo {volume}
  {4}},\ \bibinfo {pages} {033034} (\bibinfo {year}
  {2022}{\natexlab{b}})}\BibitemShut {NoStop}%
\bibitem [{\citenamefont {Katabarwa}\ \emph {et~al.}(2021)\citenamefont
  {Katabarwa}, \citenamefont {Kunitsa}, \citenamefont {Peropadre},\ and\
  \citenamefont {Johnson}}]{katabarwa_reducing_2021}%
  \BibitemOpen
  \bibfield  {author} {\bibinfo {author} {\bibfnamefont {A.}~\bibnamefont
  {Katabarwa}}, \bibinfo {author} {\bibfnamefont {A.}~\bibnamefont {Kunitsa}},
  \bibinfo {author} {\bibfnamefont {B.}~\bibnamefont {Peropadre}},\ and\
  \bibinfo {author} {\bibfnamefont {P.}~\bibnamefont {Johnson}},\ }\href
  {https://doi.org/10.48550/arXiv.2110.10664} {\bibinfo {title} {Reducing
  runtime and error in {VQE} using deeper and noisier quantum circuits}}
  (\bibinfo {year} {2021}),\ \bibinfo {note} {arXiv:2110.10664
  [quant-ph]}\BibitemShut {NoStop}%
\bibitem [{\citenamefont {Dalal}\ and\ \citenamefont
  {Katabarwa}(2023)}]{dalal_noise_2023}%
  \BibitemOpen
  \bibfield  {author} {\bibinfo {author} {\bibfnamefont {A.}~\bibnamefont
  {Dalal}}\ and\ \bibinfo {author} {\bibfnamefont {A.}~\bibnamefont
  {Katabarwa}},\ }\bibfield  {title} {\bibinfo {title} {Noise tailoring for
  robust amplitude estimation},\ }\href
  {https://doi.org/10.1088/1367-2630/acb5bc} {\bibfield  {journal} {\bibinfo
  {journal} {New Journal of Physics}\ }\textbf {\bibinfo {volume} {25}},\
  \bibinfo {pages} {023015} (\bibinfo {year} {2023})}\BibitemShut {NoStop}%
\bibitem [{\citenamefont {Grinko}\ \emph {et~al.}(2021)\citenamefont {Grinko},
  \citenamefont {Gacon}, \citenamefont {Zoufal},\ and\ \citenamefont
  {Woerner}}]{grinko_iterative_2021}%
  \BibitemOpen
  \bibfield  {author} {\bibinfo {author} {\bibfnamefont {D.}~\bibnamefont
  {Grinko}}, \bibinfo {author} {\bibfnamefont {J.}~\bibnamefont {Gacon}},
  \bibinfo {author} {\bibfnamefont {C.}~\bibnamefont {Zoufal}},\ and\ \bibinfo
  {author} {\bibfnamefont {S.}~\bibnamefont {Woerner}},\ }\bibfield  {title}
  {\bibinfo {title} {Iterative {Quantum} {Amplitude} {Estimation}},\ }\href
  {https://doi.org/10.1038/s41534-021-00379-1} {\bibfield  {journal} {\bibinfo
  {journal} {npj Quantum Information}\ }\textbf {\bibinfo {volume} {7}},\
  \bibinfo {pages} {52} (\bibinfo {year} {2021})},\ \bibinfo {note}
  {arXiv:1912.05559 [quant-ph]}\BibitemShut {NoStop}%
\bibitem [{\citenamefont {Rao}\ \emph {et~al.}(2020)\citenamefont {Rao},
  \citenamefont {Yu}, \citenamefont {Lim}, \citenamefont {Jin},\ and\
  \citenamefont {Choi}}]{rao2020quantum}%
  \BibitemOpen
  \bibfield  {author} {\bibinfo {author} {\bibfnamefont {P.}~\bibnamefont
  {Rao}}, \bibinfo {author} {\bibfnamefont {K.}~\bibnamefont {Yu}}, \bibinfo
  {author} {\bibfnamefont {H.}~\bibnamefont {Lim}}, \bibinfo {author}
  {\bibfnamefont {D.}~\bibnamefont {Jin}},\ and\ \bibinfo {author}
  {\bibfnamefont {D.}~\bibnamefont {Choi}},\ }\bibfield  {title} {\bibinfo
  {title} {Quantum amplitude estimation algorithms on ibm quantum devices},\
  }in\ \href@noop {} {\emph {\bibinfo {booktitle} {Quantum Communications and
  Quantum Imaging XVIII}}},\ Vol.\ \bibinfo {volume} {11507}\ (\bibinfo
  {organization} {SPIE},\ \bibinfo {year} {2020})\ pp.\ \bibinfo {pages}
  {49--60}\BibitemShut {NoStop}%
\bibitem [{\citenamefont {Simon}\ \emph {et~al.}(2024)\citenamefont {Simon},
  \citenamefont {Degroote}, \citenamefont {Moll}, \citenamefont {Santagati},
  \citenamefont {Streif},\ and\ \citenamefont {Wiebe}}]{simon2024amplified}%
  \BibitemOpen
  \bibfield  {author} {\bibinfo {author} {\bibfnamefont {S.}~\bibnamefont
  {Simon}}, \bibinfo {author} {\bibfnamefont {M.}~\bibnamefont {Degroote}},
  \bibinfo {author} {\bibfnamefont {N.}~\bibnamefont {Moll}}, \bibinfo {author}
  {\bibfnamefont {R.}~\bibnamefont {Santagati}}, \bibinfo {author}
  {\bibfnamefont {M.}~\bibnamefont {Streif}},\ and\ \bibinfo {author}
  {\bibfnamefont {N.}~\bibnamefont {Wiebe}},\ }\bibfield  {title} {\bibinfo
  {title} {Amplified amplitude estimation: Exploiting prior knowledge to
  improve estimates of expectation values},\ }\href@noop {} {\bibfield
  {journal} {\bibinfo  {journal} {arXiv preprint arXiv:2402.14791}\ } (\bibinfo
  {year} {2024})}\BibitemShut {NoStop}%
\bibitem [{\citenamefont {Dasgupta}\ and\ \citenamefont
  {Humble}(2021)}]{Dasgupta2021}%
  \BibitemOpen
  \bibfield  {author} {\bibinfo {author} {\bibfnamefont {S.}~\bibnamefont
  {Dasgupta}}\ and\ \bibinfo {author} {\bibfnamefont {T.~S.}\ \bibnamefont
  {Humble}},\ }\bibfield  {title} {\bibinfo {title} {{Stability of noisy
  quantum computing devices}},\ }\href {https://arxiv.org/abs/2105.09472v1}
  {\bibfield  {journal} {\bibinfo  {journal} {arXiv preprint arXiv:2105.09472}\
  ,\ \bibinfo {pages} {1}} (\bibinfo {year} {2021})},\ \Eprint
  {https://arxiv.org/abs/2105.09472} {arXiv:2105.09472} \BibitemShut {NoStop}%
\bibitem [{\citenamefont {Tanaka}\ \emph {et~al.}(2022)\citenamefont {Tanaka},
  \citenamefont {Uno}, \citenamefont {Onodera}, \citenamefont {Yamamoto},\ and\
  \citenamefont {Suzuki}}]{tanaka_noisy_2022}%
  \BibitemOpen
  \bibfield  {author} {\bibinfo {author} {\bibfnamefont {T.}~\bibnamefont
  {Tanaka}}, \bibinfo {author} {\bibfnamefont {S.}~\bibnamefont {Uno}},
  \bibinfo {author} {\bibfnamefont {T.}~\bibnamefont {Onodera}}, \bibinfo
  {author} {\bibfnamefont {N.}~\bibnamefont {Yamamoto}},\ and\ \bibinfo
  {author} {\bibfnamefont {Y.}~\bibnamefont {Suzuki}},\ }\bibfield  {title}
  {\bibinfo {title} {Noisy quantum amplitude estimation without noise
  estimation},\ }\href {https://doi.org/10.1103/PhysRevA.105.012411} {\bibfield
   {journal} {\bibinfo  {journal} {Physical Review A}\ }\textbf {\bibinfo
  {volume} {105}},\ \bibinfo {pages} {012411} (\bibinfo {year} {2022})},\
  \bibinfo {note} {arXiv:2110.04258 [quant-ph]}\BibitemShut {NoStop}%
\bibitem [{\citenamefont {Hempel}\ \emph {et~al.}(2018)\citenamefont {Hempel},
  \citenamefont {Maier}, \citenamefont {Romero}, \citenamefont {McClean},
  \citenamefont {Monz}, \citenamefont {Shen}, \citenamefont {Jurcevic},
  \citenamefont {Lanyon}, \citenamefont {Love}, \citenamefont {Babbush},
  \citenamefont {Aspuru-Guzik}, \citenamefont {Blatt},\ and\ \citenamefont
  {Roos}}]{Hempel2018}%
  \BibitemOpen
  \bibfield  {author} {\bibinfo {author} {\bibfnamefont {C.}~\bibnamefont
  {Hempel}}, \bibinfo {author} {\bibfnamefont {C.}~\bibnamefont {Maier}},
  \bibinfo {author} {\bibfnamefont {J.}~\bibnamefont {Romero}}, \bibinfo
  {author} {\bibfnamefont {J.}~\bibnamefont {McClean}}, \bibinfo {author}
  {\bibfnamefont {T.}~\bibnamefont {Monz}}, \bibinfo {author} {\bibfnamefont
  {H.}~\bibnamefont {Shen}}, \bibinfo {author} {\bibfnamefont {P.}~\bibnamefont
  {Jurcevic}}, \bibinfo {author} {\bibfnamefont {B.~P.}\ \bibnamefont
  {Lanyon}}, \bibinfo {author} {\bibfnamefont {P.}~\bibnamefont {Love}},
  \bibinfo {author} {\bibfnamefont {R.}~\bibnamefont {Babbush}}, \bibinfo
  {author} {\bibfnamefont {A.}~\bibnamefont {Aspuru-Guzik}}, \bibinfo {author}
  {\bibfnamefont {R.}~\bibnamefont {Blatt}},\ and\ \bibinfo {author}
  {\bibfnamefont {C.~F.}\ \bibnamefont {Roos}},\ }\bibfield  {title} {\bibinfo
  {title} {{Quantum Chemistry Calculations on a Trapped-Ion Quantum
  Simulator}},\ }\href {https://doi.org/10.1103/PhysRevX.8.031022} {\bibfield
  {journal} {\bibinfo  {journal} {Physical Review X}\ }\textbf {\bibinfo
  {volume} {8}},\ \bibinfo {pages} {031022} (\bibinfo {year}
  {2018})}\BibitemShut {NoStop}%
\bibitem [{\citenamefont {Romero}\ \emph {et~al.}(2018)\citenamefont {Romero},
  \citenamefont {Babbush}, \citenamefont {McClean}, \citenamefont {Hempel},
  \citenamefont {Love},\ and\ \citenamefont
  {Aspuru-Guzik}}]{romero_strategies_2018}%
  \BibitemOpen
  \bibfield  {author} {\bibinfo {author} {\bibfnamefont {J.}~\bibnamefont
  {Romero}}, \bibinfo {author} {\bibfnamefont {R.}~\bibnamefont {Babbush}},
  \bibinfo {author} {\bibfnamefont {J.~R.}\ \bibnamefont {McClean}}, \bibinfo
  {author} {\bibfnamefont {C.}~\bibnamefont {Hempel}}, \bibinfo {author}
  {\bibfnamefont {P.}~\bibnamefont {Love}},\ and\ \bibinfo {author}
  {\bibfnamefont {A.}~\bibnamefont {Aspuru-Guzik}},\ }\href
  {https://doi.org/10.48550/arXiv.1701.02691} {\bibinfo {title} {Strategies for
  quantum computing molecular energies using the unitary coupled cluster
  ansatz}} (\bibinfo {year} {2018}),\ \bibinfo {note} {arXiv:1701.02691
  [quant-ph]}\BibitemShut {NoStop}%
\bibitem [{\citenamefont {Zhao}\ \emph {et~al.}(2020)\citenamefont {Zhao},
  \citenamefont {Tranter}, \citenamefont {Kirby}, \citenamefont {Ung},
  \citenamefont {Miyake},\ and\ \citenamefont {Love}}]{zhao_measurement_2020}%
  \BibitemOpen
  \bibfield  {author} {\bibinfo {author} {\bibfnamefont {A.}~\bibnamefont
  {Zhao}}, \bibinfo {author} {\bibfnamefont {A.}~\bibnamefont {Tranter}},
  \bibinfo {author} {\bibfnamefont {W.~M.}\ \bibnamefont {Kirby}}, \bibinfo
  {author} {\bibfnamefont {S.~F.}\ \bibnamefont {Ung}}, \bibinfo {author}
  {\bibfnamefont {A.}~\bibnamefont {Miyake}},\ and\ \bibinfo {author}
  {\bibfnamefont {P.~J.}\ \bibnamefont {Love}},\ }\bibfield  {title} {\bibinfo
  {title} {Measurement reduction in variational quantum algorithms},\ }\href
  {https://doi.org/10.1103/PhysRevA.101.062322} {\bibfield  {journal} {\bibinfo
   {journal} {Physical Review A}\ }\textbf {\bibinfo {volume} {101}},\ \bibinfo
  {pages} {062322} (\bibinfo {year} {2020})}\BibitemShut {NoStop}%
\bibitem [{\citenamefont {Verteletskyi}\ \emph {et~al.}()\citenamefont
  {Verteletskyi}, \citenamefont {Yen},\ and\ \citenamefont
  {Izmaylov}}]{min_clique_grouping}%
  \BibitemOpen
  \bibfield  {author} {\bibinfo {author} {\bibfnamefont {V.}~\bibnamefont
  {Verteletskyi}}, \bibinfo {author} {\bibfnamefont {T.-C.}\ \bibnamefont
  {Yen}},\ and\ \bibinfo {author} {\bibfnamefont {A.~F.}\ \bibnamefont
  {Izmaylov}},\ }\href@noop {} {\bibinfo {title} {Measurement optimization in
  the variational quantum eigensolver using a minimum clique cover}},\ \Eprint
  {https://arxiv.org/abs/arXiv:1907.03358v4} {arXiv:1907.03358v4} \BibitemShut
  {NoStop}%
\bibitem [{\citenamefont {{Gokhale}}\ \emph {et~al.}(2019)\citenamefont
  {{Gokhale}}, \citenamefont {{Angiuli}}, \citenamefont {{Ding}}, \citenamefont
  {{Gui}}, \citenamefont {{Tomesh}}, \citenamefont {{Suchara}}, \citenamefont
  {{Martonosi}},\ and\ \citenamefont {{Chong}}}]{linear_grouping_algo}%
  \BibitemOpen
  \bibfield  {author} {\bibinfo {author} {\bibfnamefont {P.}~\bibnamefont
  {{Gokhale}}}, \bibinfo {author} {\bibfnamefont {O.}~\bibnamefont
  {{Angiuli}}}, \bibinfo {author} {\bibfnamefont {Y.}~\bibnamefont {{Ding}}},
  \bibinfo {author} {\bibfnamefont {K.}~\bibnamefont {{Gui}}}, \bibinfo
  {author} {\bibfnamefont {T.}~\bibnamefont {{Tomesh}}}, \bibinfo {author}
  {\bibfnamefont {M.}~\bibnamefont {{Suchara}}}, \bibinfo {author}
  {\bibfnamefont {M.}~\bibnamefont {{Martonosi}}},\ and\ \bibinfo {author}
  {\bibfnamefont {F.~T.}\ \bibnamefont {{Chong}}},\ }\href
  {https://ui.adsabs.harvard.edu/abs/2019arXiv190713623G} {\bibinfo {title}
  {{Minimizing State Preparations in Variational Quantum Eigensolver by
  Partitioning into Commuting Families}}} (\bibinfo {year} {2019}),\ \Eprint
  {https://arxiv.org/abs/1907.13623} {arXiv:1907.13623} \BibitemShut {NoStop}%
\bibitem [{\citenamefont {Yen}\ and\ \citenamefont
  {Izmaylov}(2021)}]{yen2021cartan}%
  \BibitemOpen
  \bibfield  {author} {\bibinfo {author} {\bibfnamefont {T.-C.}\ \bibnamefont
  {Yen}}\ and\ \bibinfo {author} {\bibfnamefont {A.~F.}\ \bibnamefont
  {Izmaylov}},\ }\bibfield  {title} {\bibinfo {title} {Cartan subalgebra
  approach to efficient measurements of quantum observables},\ }\href@noop {}
  {\bibfield  {journal} {\bibinfo  {journal} {PRX quantum}\ }\textbf {\bibinfo
  {volume} {2}},\ \bibinfo {pages} {040320} (\bibinfo {year}
  {2021})}\BibitemShut {NoStop}%
\bibitem [{\citenamefont {Izmaylov}\ \emph {et~al.}(2019)\citenamefont
  {Izmaylov}, \citenamefont {Yen}, \citenamefont {Lang},\ and\ \citenamefont
  {Verteletskyi}}]{izmaylov_unitary_2019}%
  \BibitemOpen
  \bibfield  {author} {\bibinfo {author} {\bibfnamefont {A.~F.}\ \bibnamefont
  {Izmaylov}}, \bibinfo {author} {\bibfnamefont {T.-C.}\ \bibnamefont {Yen}},
  \bibinfo {author} {\bibfnamefont {R.~A.}\ \bibnamefont {Lang}},\ and\
  \bibinfo {author} {\bibfnamefont {V.}~\bibnamefont {Verteletskyi}},\ }\href
  {https://doi.org/10.48550/arXiv.1907.09040} {\bibinfo {title} {Unitary
  partitioning approach to the measurement problem in the {Variational}
  {Quantum} {Eigensolver} method}} (\bibinfo {year} {2019})\BibitemShut
  {NoStop}%
\bibitem [{\citenamefont {Wasserman}(2004)}]{wasserman_all_2004}%
  \BibitemOpen
  \bibfield  {author} {\bibinfo {author} {\bibfnamefont {L.}~\bibnamefont
  {Wasserman}},\ }\href {https://doi.org/10.1007/978-0-387-21736-9} {\emph
  {\bibinfo {title} {All of {Statistics}: {A} {Concise} {Course} in
  {Statistical} {Inference}}}},\ Springer {Texts} in {Statistics}\ (\bibinfo
  {publisher} {Springer},\ \bibinfo {address} {New York, NY},\ \bibinfo {year}
  {2004})\BibitemShut {NoStop}%
\bibitem [{\citenamefont {Brown}\ \emph {et~al.}(2020)\citenamefont {Brown},
  \citenamefont {Goktas},\ and\ \citenamefont {Tham}}]{brown_quantum_2020}%
  \BibitemOpen
  \bibfield  {author} {\bibinfo {author} {\bibfnamefont {E.~G.}\ \bibnamefont
  {Brown}}, \bibinfo {author} {\bibfnamefont {O.}~\bibnamefont {Goktas}},\ and\
  \bibinfo {author} {\bibfnamefont {W.~K.}\ \bibnamefont {Tham}},\ }\href
  {https://doi.org/10.48550/arXiv.2006.14145} {\bibinfo {title} {Quantum
  {Amplitude} {Estimation} in the {Presence} of {Noise}}} (\bibinfo {year}
  {2020}),\ \bibinfo {note} {arXiv:2006.14145 [quant-ph]}\BibitemShut {NoStop}%
\bibitem [{\citenamefont {Callison}\ and\ \citenamefont
  {Browne}(2023)}]{callison_improved_2023}%
  \BibitemOpen
  \bibfield  {author} {\bibinfo {author} {\bibfnamefont {A.}~\bibnamefont
  {Callison}}\ and\ \bibinfo {author} {\bibfnamefont {D.~E.}\ \bibnamefont
  {Browne}},\ }\href {https://doi.org/10.48550/arXiv.2209.03321} {\bibinfo
  {title} {Improved maximum-likelihood quantum amplitude estimation}} (\bibinfo
  {year} {2023}),\ \bibinfo {note} {arXiv:2209.03321 [quant-ph]}\BibitemShut
  {NoStop}%
\bibitem [{\citenamefont {Cross}\ \emph {et~al.}(2019)\citenamefont {Cross},
  \citenamefont {Bishop}, \citenamefont {Sheldon}, \citenamefont {Nation},\
  and\ \citenamefont {Gambetta}}]{cross_validating_2019}%
  \BibitemOpen
  \bibfield  {author} {\bibinfo {author} {\bibfnamefont {A.~W.}\ \bibnamefont
  {Cross}}, \bibinfo {author} {\bibfnamefont {L.~S.}\ \bibnamefont {Bishop}},
  \bibinfo {author} {\bibfnamefont {S.}~\bibnamefont {Sheldon}}, \bibinfo
  {author} {\bibfnamefont {P.~D.}\ \bibnamefont {Nation}},\ and\ \bibinfo
  {author} {\bibfnamefont {J.~M.}\ \bibnamefont {Gambetta}},\ }\bibfield
  {title} {\bibinfo {title} {Validating quantum computers using randomized
  model circuits},\ }\href {https://doi.org/10.1103/PhysRevA.100.032328}
  {\bibfield  {journal} {\bibinfo  {journal} {Physical Review A}\ }\textbf
  {\bibinfo {volume} {100}},\ \bibinfo {pages} {032328} (\bibinfo {year}
  {2019})},\ \bibinfo {note} {publisher: American Physical Society}\BibitemShut
  {NoStop}%
\bibitem [{\citenamefont {Blume-Kohout}\ and\ \citenamefont
  {Young}(2020)}]{blume-kohout_volumetric_2020}%
  \BibitemOpen
  \bibfield  {author} {\bibinfo {author} {\bibfnamefont {R.}~\bibnamefont
  {Blume-Kohout}}\ and\ \bibinfo {author} {\bibfnamefont {K.~C.}\ \bibnamefont
  {Young}},\ }\bibfield  {title} {\bibinfo {title} {A volumetric framework for
  quantum computer benchmarks},\ }\href
  {https://doi.org/10.22331/q-2020-11-15-362} {\bibfield  {journal} {\bibinfo
  {journal} {Quantum}\ }\textbf {\bibinfo {volume} {4}},\ \bibinfo {pages}
  {362} (\bibinfo {year} {2020})},\ \bibinfo {note} {publisher: Verein zur
  Förderung des Open Access Publizierens in den
  Quantenwissenschaften}\BibitemShut {NoStop}%
\bibitem [{\citenamefont {Bravyi}\ \emph {et~al.}(2021)\citenamefont {Bravyi},
  \citenamefont {Sheldon}, \citenamefont {Kandala}, \citenamefont {Mckay},\
  and\ \citenamefont {Gambetta}}]{bravyi_mitigating_2021}%
  \BibitemOpen
  \bibfield  {author} {\bibinfo {author} {\bibfnamefont {S.}~\bibnamefont
  {Bravyi}}, \bibinfo {author} {\bibfnamefont {S.}~\bibnamefont {Sheldon}},
  \bibinfo {author} {\bibfnamefont {A.}~\bibnamefont {Kandala}}, \bibinfo
  {author} {\bibfnamefont {D.~C.}\ \bibnamefont {Mckay}},\ and\ \bibinfo
  {author} {\bibfnamefont {J.~M.}\ \bibnamefont {Gambetta}},\ }\bibfield
  {title} {\bibinfo {title} {Mitigating measurement errors in multi-qubit
  experiments},\ }\bibfield  {journal} {\bibinfo  {journal} {Physical Review
  A}\ }\textbf {\bibinfo {volume} {103}},\ \href
  {https://doi.org/10.1103/PhysRevA.103.042605} {10.1103/PhysRevA.103.042605}
  (\bibinfo {year} {2021})\BibitemShut {NoStop}%
\bibitem [{\citenamefont {Virtanen}\ \emph {et~al.}(2020)\citenamefont
  {Virtanen}, \citenamefont {Gommers}, \citenamefont {Oliphant}, \citenamefont
  {Haberland}, \citenamefont {Reddy}, \citenamefont {Cournapeau}, \citenamefont
  {Burovski}, \citenamefont {Peterson}, \citenamefont {Weckesser},
  \citenamefont {Bright}, \citenamefont {{van der Walt}}, \citenamefont
  {Brett}, \citenamefont {Wilson}, \citenamefont {Millman}, \citenamefont
  {Mayorov}, \citenamefont {Nelson}, \citenamefont {Jones}, \citenamefont
  {Kern}, \citenamefont {Larson}, \citenamefont {Carey}, \citenamefont {Polat},
  \citenamefont {Feng}, \citenamefont {Moore}, \citenamefont {{VanderPlas}},
  \citenamefont {Laxalde}, \citenamefont {Perktold}, \citenamefont {Cimrman},
  \citenamefont {Henriksen}, \citenamefont {Quintero}, \citenamefont {Harris},
  \citenamefont {Archibald}, \citenamefont {Ribeiro}, \citenamefont
  {Pedregosa}, \citenamefont {{van Mulbregt}},\ and\ \citenamefont {{SciPy 1.0
  Contributors}}}]{2020SciPy-NMeth}%
  \BibitemOpen
  \bibfield  {author} {\bibinfo {author} {\bibfnamefont {P.}~\bibnamefont
  {Virtanen}}, \bibinfo {author} {\bibfnamefont {R.}~\bibnamefont {Gommers}},
  \bibinfo {author} {\bibfnamefont {T.~E.}\ \bibnamefont {Oliphant}}, \bibinfo
  {author} {\bibfnamefont {M.}~\bibnamefont {Haberland}}, \bibinfo {author}
  {\bibfnamefont {T.}~\bibnamefont {Reddy}}, \bibinfo {author} {\bibfnamefont
  {D.}~\bibnamefont {Cournapeau}}, \bibinfo {author} {\bibfnamefont
  {E.}~\bibnamefont {Burovski}}, \bibinfo {author} {\bibfnamefont
  {P.}~\bibnamefont {Peterson}}, \bibinfo {author} {\bibfnamefont
  {W.}~\bibnamefont {Weckesser}}, \bibinfo {author} {\bibfnamefont
  {J.}~\bibnamefont {Bright}}, \bibinfo {author} {\bibfnamefont {S.~J.}\
  \bibnamefont {{van der Walt}}}, \bibinfo {author} {\bibfnamefont
  {M.}~\bibnamefont {Brett}}, \bibinfo {author} {\bibfnamefont
  {J.}~\bibnamefont {Wilson}}, \bibinfo {author} {\bibfnamefont {K.~J.}\
  \bibnamefont {Millman}}, \bibinfo {author} {\bibfnamefont {N.}~\bibnamefont
  {Mayorov}}, \bibinfo {author} {\bibfnamefont {A.~R.~J.}\ \bibnamefont
  {Nelson}}, \bibinfo {author} {\bibfnamefont {E.}~\bibnamefont {Jones}},
  \bibinfo {author} {\bibfnamefont {R.}~\bibnamefont {Kern}}, \bibinfo {author}
  {\bibfnamefont {E.}~\bibnamefont {Larson}}, \bibinfo {author} {\bibfnamefont
  {C.~J.}\ \bibnamefont {Carey}}, \bibinfo {author} {\bibfnamefont
  {{\.I}.}~\bibnamefont {Polat}}, \bibinfo {author} {\bibfnamefont
  {Y.}~\bibnamefont {Feng}}, \bibinfo {author} {\bibfnamefont {E.~W.}\
  \bibnamefont {Moore}}, \bibinfo {author} {\bibfnamefont {J.}~\bibnamefont
  {{VanderPlas}}}, \bibinfo {author} {\bibfnamefont {D.}~\bibnamefont
  {Laxalde}}, \bibinfo {author} {\bibfnamefont {J.}~\bibnamefont {Perktold}},
  \bibinfo {author} {\bibfnamefont {R.}~\bibnamefont {Cimrman}}, \bibinfo
  {author} {\bibfnamefont {I.}~\bibnamefont {Henriksen}}, \bibinfo {author}
  {\bibfnamefont {E.~A.}\ \bibnamefont {Quintero}}, \bibinfo {author}
  {\bibfnamefont {C.~R.}\ \bibnamefont {Harris}}, \bibinfo {author}
  {\bibfnamefont {A.~M.}\ \bibnamefont {Archibald}}, \bibinfo {author}
  {\bibfnamefont {A.~H.}\ \bibnamefont {Ribeiro}}, \bibinfo {author}
  {\bibfnamefont {F.}~\bibnamefont {Pedregosa}}, \bibinfo {author}
  {\bibfnamefont {P.}~\bibnamefont {{van Mulbregt}}},\ and\ \bibinfo {author}
  {\bibnamefont {{SciPy 1.0 Contributors}}},\ }\bibfield  {title} {\bibinfo
  {title} {{{SciPy} 1.0: Fundamental Algorithms for Scientific Computing in
  Python}},\ }\href {https://doi.org/10.1038/s41592-019-0686-2} {\bibfield
  {journal} {\bibinfo  {journal} {Nature Methods}\ }\textbf {\bibinfo {volume}
  {17}},\ \bibinfo {pages} {261} (\bibinfo {year} {2020})}\BibitemShut
  {NoStop}%
\bibitem [{\citenamefont {Efron}\ and\ \citenamefont
  {Tibshirani}(1994)}]{efron_introduction_1994}%
  \BibitemOpen
  \bibfield  {author} {\bibinfo {author} {\bibfnamefont {B.}~\bibnamefont
  {Efron}}\ and\ \bibinfo {author} {\bibfnamefont {R.~J.}\ \bibnamefont
  {Tibshirani}},\ }\href {https://doi.org/10.1201/9780429246593} {\emph
  {\bibinfo {title} {An {Introduction} to the {Bootstrap}}}}\ (\bibinfo
  {publisher} {Chapman and Hall/CRC},\ \bibinfo {address} {New York},\ \bibinfo
  {year} {1994})\BibitemShut {NoStop}%
\bibitem [{\citenamefont {Javadi-Abhari}\ \emph {et~al.}(2024)\citenamefont
  {Javadi-Abhari}, \citenamefont {Treinish}, \citenamefont {Krsulich},
  \citenamefont {Wood}, \citenamefont {Lishman}, \citenamefont {Gacon},
  \citenamefont {Martiel}, \citenamefont {Nation}, \citenamefont {Bishop},
  \citenamefont {Cross}, \citenamefont {Johnson},\ and\ \citenamefont
  {Gambetta}}]{javadi-abhari_quantum_2024}%
  \BibitemOpen
  \bibfield  {author} {\bibinfo {author} {\bibfnamefont {A.}~\bibnamefont
  {Javadi-Abhari}}, \bibinfo {author} {\bibfnamefont {M.}~\bibnamefont
  {Treinish}}, \bibinfo {author} {\bibfnamefont {K.}~\bibnamefont {Krsulich}},
  \bibinfo {author} {\bibfnamefont {C.~J.}\ \bibnamefont {Wood}}, \bibinfo
  {author} {\bibfnamefont {J.}~\bibnamefont {Lishman}}, \bibinfo {author}
  {\bibfnamefont {J.}~\bibnamefont {Gacon}}, \bibinfo {author} {\bibfnamefont
  {S.}~\bibnamefont {Martiel}}, \bibinfo {author} {\bibfnamefont {P.~D.}\
  \bibnamefont {Nation}}, \bibinfo {author} {\bibfnamefont {L.~S.}\
  \bibnamefont {Bishop}}, \bibinfo {author} {\bibfnamefont {A.~W.}\
  \bibnamefont {Cross}}, \bibinfo {author} {\bibfnamefont {B.~R.}\ \bibnamefont
  {Johnson}},\ and\ \bibinfo {author} {\bibfnamefont {J.~M.}\ \bibnamefont
  {Gambetta}},\ }\href {https://doi.org/10.48550/arXiv.2405.08810} {\bibinfo
  {title} {Quantum computing with {Qiskit}}} (\bibinfo {year} {2024}),\
  \bibinfo {note} {arXiv:2405.08810 [quant-ph]}\BibitemShut {NoStop}%
\bibitem [{\citenamefont {Temme}\ \emph {et~al.}(2017)\citenamefont {Temme},
  \citenamefont {Bravyi},\ and\ \citenamefont {Gambetta}}]{temme_error_2017}%
  \BibitemOpen
  \bibfield  {author} {\bibinfo {author} {\bibfnamefont {K.}~\bibnamefont
  {Temme}}, \bibinfo {author} {\bibfnamefont {S.}~\bibnamefont {Bravyi}},\ and\
  \bibinfo {author} {\bibfnamefont {J.~M.}\ \bibnamefont {Gambetta}},\
  }\bibfield  {title} {\bibinfo {title} {Error {Mitigation} for {Short}-{Depth}
  {Quantum} {Circuits}},\ }\href
  {https://doi.org/10.1103/PhysRevLett.119.180509} {\bibfield  {journal}
  {\bibinfo  {journal} {Physical Review Letters}\ }\textbf {\bibinfo {volume}
  {119}},\ \bibinfo {pages} {180509} (\bibinfo {year} {2017})},\ \bibinfo
  {note} {publisher: American Physical Society}\BibitemShut {NoStop}%
\bibitem [{\citenamefont {Kandala}\ \emph {et~al.}(2019)\citenamefont
  {Kandala}, \citenamefont {Temme}, \citenamefont {Córcoles}, \citenamefont
  {Mezzacapo}, \citenamefont {Chow},\ and\ \citenamefont
  {Gambetta}}]{kandala_error_2019}%
  \BibitemOpen
  \bibfield  {author} {\bibinfo {author} {\bibfnamefont {A.}~\bibnamefont
  {Kandala}}, \bibinfo {author} {\bibfnamefont {K.}~\bibnamefont {Temme}},
  \bibinfo {author} {\bibfnamefont {A.~D.}\ \bibnamefont {Córcoles}}, \bibinfo
  {author} {\bibfnamefont {A.}~\bibnamefont {Mezzacapo}}, \bibinfo {author}
  {\bibfnamefont {J.~M.}\ \bibnamefont {Chow}},\ and\ \bibinfo {author}
  {\bibfnamefont {J.~M.}\ \bibnamefont {Gambetta}},\ }\bibfield  {title}
  {\bibinfo {title} {Error mitigation extends the computational reach of a
  noisy quantum processor},\ }\href {https://doi.org/10.1038/s41586-019-1040-7}
  {\bibfield  {journal} {\bibinfo  {journal} {Nature}\ }\textbf {\bibinfo
  {volume} {567}},\ \bibinfo {pages} {491} (\bibinfo {year} {2019})},\ \bibinfo
  {note} {publisher: Nature Publishing Group}\BibitemShut {NoStop}%
\bibitem [{\citenamefont {Krebsbach}\ \emph {et~al.}(2022)\citenamefont
  {Krebsbach}, \citenamefont {Trauzettel},\ and\ \citenamefont
  {Calzona}}]{krebsbach_optimization_2022}%
  \BibitemOpen
  \bibfield  {author} {\bibinfo {author} {\bibfnamefont {M.}~\bibnamefont
  {Krebsbach}}, \bibinfo {author} {\bibfnamefont {B.}~\bibnamefont
  {Trauzettel}},\ and\ \bibinfo {author} {\bibfnamefont {A.}~\bibnamefont
  {Calzona}},\ }\bibfield  {title} {\bibinfo {title} {Optimization of
  {Richardson} extrapolation for quantum error mitigation},\ }\href
  {https://doi.org/10.1103/PhysRevA.106.062436} {\bibfield  {journal} {\bibinfo
   {journal} {Physical Review A}\ }\textbf {\bibinfo {volume} {106}},\ \bibinfo
  {pages} {062436} (\bibinfo {year} {2022})},\ \bibinfo {note}
  {arXiv:2201.08080 [cond-mat, physics:quant-ph]}\BibitemShut {NoStop}%
\bibitem [{\citenamefont {Czarnik}\ \emph {et~al.}(2021)\citenamefont
  {Czarnik}, \citenamefont {Arrasmith}, \citenamefont {Coles},\ and\
  \citenamefont {Cincio}}]{czarnik_error_2021}%
  \BibitemOpen
  \bibfield  {author} {\bibinfo {author} {\bibfnamefont {P.}~\bibnamefont
  {Czarnik}}, \bibinfo {author} {\bibfnamefont {A.}~\bibnamefont {Arrasmith}},
  \bibinfo {author} {\bibfnamefont {P.~J.}\ \bibnamefont {Coles}},\ and\
  \bibinfo {author} {\bibfnamefont {L.}~\bibnamefont {Cincio}},\ }\bibfield
  {title} {\bibinfo {title} {Error mitigation with {Clifford} quantum-circuit
  data},\ }\href {https://doi.org/10.22331/q-2021-11-26-592} {\bibfield
  {journal} {\bibinfo  {journal} {Quantum}\ }\textbf {\bibinfo {volume} {5}},\
  \bibinfo {pages} {592} (\bibinfo {year} {2021})},\ \bibinfo {note}
  {arXiv:2005.10189 [quant-ph]}\BibitemShut {NoStop}%
\bibitem [{\citenamefont {Lowe}\ \emph {et~al.}(2021)\citenamefont {Lowe},
  \citenamefont {Gordon}, \citenamefont {Czarnik}, \citenamefont {Arrasmith},
  \citenamefont {Coles},\ and\ \citenamefont {Cincio}}]{lowe_unified_2021}%
  \BibitemOpen
  \bibfield  {author} {\bibinfo {author} {\bibfnamefont {A.}~\bibnamefont
  {Lowe}}, \bibinfo {author} {\bibfnamefont {M.~H.}\ \bibnamefont {Gordon}},
  \bibinfo {author} {\bibfnamefont {P.}~\bibnamefont {Czarnik}}, \bibinfo
  {author} {\bibfnamefont {A.}~\bibnamefont {Arrasmith}}, \bibinfo {author}
  {\bibfnamefont {P.~J.}\ \bibnamefont {Coles}},\ and\ \bibinfo {author}
  {\bibfnamefont {L.}~\bibnamefont {Cincio}},\ }\bibfield  {title} {\bibinfo
  {title} {Unified approach to data-driven quantum error mitigation},\ }\href
  {https://doi.org/10.1103/PhysRevResearch.3.033098} {\bibfield  {journal}
  {\bibinfo  {journal} {Physical Review Research}\ }\textbf {\bibinfo {volume}
  {3}},\ \bibinfo {pages} {033098} (\bibinfo {year} {2021})},\ \bibinfo {note}
  {arXiv:2011.01157 [quant-ph]}\BibitemShut {NoStop}%
\bibitem [{\citenamefont {Saki}\ \emph {et~al.}(2023)\citenamefont {Saki},
  \citenamefont {Katabarwa}, \citenamefont {Resch},\ and\ \citenamefont
  {Umbrarescu}}]{saki_hypothesis_2023}%
  \BibitemOpen
  \bibfield  {author} {\bibinfo {author} {\bibfnamefont {A.~A.}\ \bibnamefont
  {Saki}}, \bibinfo {author} {\bibfnamefont {A.}~\bibnamefont {Katabarwa}},
  \bibinfo {author} {\bibfnamefont {S.}~\bibnamefont {Resch}},\ and\ \bibinfo
  {author} {\bibfnamefont {G.}~\bibnamefont {Umbrarescu}},\ }\href
  {https://doi.org/10.48550/arXiv.2301.02690} {\bibinfo {title} {Hypothesis
  {Testing} for {Error} {Mitigation}: {How} to {Evaluate} {Error}
  {Mitigation}}} (\bibinfo {year} {2023}),\ \bibinfo {note} {arXiv:2301.02690
  [quant-ph]}\BibitemShut {NoStop}%
\bibitem [{\citenamefont {Loaiza}\ \emph {et~al.}(2023)\citenamefont {Loaiza},
  \citenamefont {Khah}, \citenamefont {Wiebe},\ and\ \citenamefont
  {Izmaylov}}]{loaiza_reducing_2023}%
  \BibitemOpen
  \bibfield  {author} {\bibinfo {author} {\bibfnamefont {I.}~\bibnamefont
  {Loaiza}}, \bibinfo {author} {\bibfnamefont {A.~M.}\ \bibnamefont {Khah}},
  \bibinfo {author} {\bibfnamefont {N.}~\bibnamefont {Wiebe}},\ and\ \bibinfo
  {author} {\bibfnamefont {A.~F.}\ \bibnamefont {Izmaylov}},\ }\bibfield
  {title} {\bibinfo {title} {Reducing molecular electronic {Hamiltonian}
  simulation cost for linear combination of unitaries approaches},\ }\href
  {https://doi.org/10.1088/2058-9565/acd577} {\bibfield  {journal} {\bibinfo
  {journal} {Quantum Science and Technology}\ }\textbf {\bibinfo {volume}
  {8}},\ \bibinfo {pages} {035019} (\bibinfo {year} {2023})}\BibitemShut
  {NoStop}%
\bibitem [{\citenamefont {Loaiza}\ and\ \citenamefont
  {Izmaylov}(2023)}]{loaiza_block-invariant_2023}%
  \BibitemOpen
  \bibfield  {author} {\bibinfo {author} {\bibfnamefont {I.}~\bibnamefont
  {Loaiza}}\ and\ \bibinfo {author} {\bibfnamefont {A.~F.}\ \bibnamefont
  {Izmaylov}},\ }\bibfield  {title} {\bibinfo {title} {Block-{Invariant}
  {Symmetry} {Shift}: {Preprocessing} {Technique} for {Second}-{Quantized}
  {Hamiltonians} to {Improve} {Their} {Decompositions} to {Linear}
  {Combination} of {Unitaries}},\ }\href
  {https://doi.org/10.1021/acs.jctc.3c00912} {\bibfield  {journal} {\bibinfo
  {journal} {Journal of Chemical Theory and Computation}\ }\textbf {\bibinfo
  {volume} {19}},\ \bibinfo {pages} {8201} (\bibinfo {year}
  {2023})}\BibitemShut {NoStop}%
\bibitem [{\citenamefont {Loaiza}\ and\ \citenamefont
  {Izmaylov}(2024)}]{loaiza_majorana_2024}%
  \BibitemOpen
  \bibfield  {author} {\bibinfo {author} {\bibfnamefont {I.}~\bibnamefont
  {Loaiza}}\ and\ \bibinfo {author} {\bibfnamefont {A.~F.}\ \bibnamefont
  {Izmaylov}},\ }\href {https://doi.org/10.48550/arXiv.2407.06571} {\bibinfo
  {title} {Majorana {Tensor} {Decomposition}: {A} unifying framework for
  decompositions of fermionic {Hamiltonians} to {Linear} {Combination} of
  {Unitaries}}} (\bibinfo {year} {2024}),\ \bibinfo {note} {arXiv:2407.06571
  [physics, physics:quant-ph]}\BibitemShut {NoStop}%
\bibitem [{\citenamefont {Rocca}\ \emph {et~al.}(2024)\citenamefont {Rocca},
  \citenamefont {Cortes}, \citenamefont {Gonthier}, \citenamefont {Ollitrault},
  \citenamefont {Parrish}, \citenamefont {Anselmetti}, \citenamefont
  {Degroote}, \citenamefont {Moll}, \citenamefont {Santagati},\ and\
  \citenamefont {Streif}}]{rocca_reducing_2024}%
  \BibitemOpen
  \bibfield  {author} {\bibinfo {author} {\bibfnamefont {D.}~\bibnamefont
  {Rocca}}, \bibinfo {author} {\bibfnamefont {C.~L.}\ \bibnamefont {Cortes}},
  \bibinfo {author} {\bibfnamefont {J.~F.}\ \bibnamefont {Gonthier}}, \bibinfo
  {author} {\bibfnamefont {P.~J.}\ \bibnamefont {Ollitrault}}, \bibinfo
  {author} {\bibfnamefont {R.~M.}\ \bibnamefont {Parrish}}, \bibinfo {author}
  {\bibfnamefont {G.-L.}\ \bibnamefont {Anselmetti}}, \bibinfo {author}
  {\bibfnamefont {M.}~\bibnamefont {Degroote}}, \bibinfo {author}
  {\bibfnamefont {N.}~\bibnamefont {Moll}}, \bibinfo {author} {\bibfnamefont
  {R.}~\bibnamefont {Santagati}},\ and\ \bibinfo {author} {\bibfnamefont
  {M.}~\bibnamefont {Streif}},\ }\bibfield  {title} {\bibinfo {title} {Reducing
  the {Runtime} of {Fault}-{Tolerant} {Quantum} {Simulations} in {Chemistry}
  through {Symmetry}-{Compressed} {Double} {Factorization}},\ }\href
  {https://doi.org/10.1021/acs.jctc.4c00352} {\bibfield  {journal} {\bibinfo
  {journal} {Journal of Chemical Theory and Computation}\ }\textbf {\bibinfo
  {volume} {20}},\ \bibinfo {pages} {4639} (\bibinfo {year}
  {2024})}\BibitemShut {NoStop}%
\bibitem [{\citenamefont {Katabarwa}\ \emph {et~al.}(2024)\citenamefont
  {Katabarwa}, \citenamefont {Gratsea}, \citenamefont {Caesura},\ and\
  \citenamefont {Johnson}}]{katabarwa_early_2024}%
  \BibitemOpen
  \bibfield  {author} {\bibinfo {author} {\bibfnamefont {A.}~\bibnamefont
  {Katabarwa}}, \bibinfo {author} {\bibfnamefont {K.}~\bibnamefont {Gratsea}},
  \bibinfo {author} {\bibfnamefont {A.}~\bibnamefont {Caesura}},\ and\ \bibinfo
  {author} {\bibfnamefont {P.~D.}\ \bibnamefont {Johnson}},\ }\bibfield
  {title} {\bibinfo {title} {Early {Fault}-{Tolerant} {Quantum} {Computing}},\
  }\href {https://doi.org/10.1103/PRXQuantum.5.020101} {\bibfield  {journal}
  {\bibinfo  {journal} {PRX Quantum}\ }\textbf {\bibinfo {volume} {5}},\
  \bibinfo {pages} {020101} (\bibinfo {year} {2024})}\BibitemShut {NoStop}%
\end{thebibliography}
\end{document}